\documentclass[3p]{elsarticle}
\usepackage{epstopdf}
\usepackage{graphicx,float}
\usepackage{amsmath, amsfonts, amssymb}
\usepackage{amsthm}
\usepackage{gensymb}
\usepackage{yhmath}
\usepackage{epsf}
\usepackage{amsfonts}
\usepackage{stmaryrd}
\usepackage{amssymb}
\usepackage{leftidx}
\usepackage{xcolor}
\usepackage{mathtools}
\usepackage{placeins}
\usepackage{booktabs}
\usepackage{enumitem}
\usepackage{caption}
\usepackage{tabu}
\usepackage{multirow}
\usepackage{stackengine}
\usepackage[utf8]{inputenc}
\usepackage[english]{babel}
\usepackage{bm}
\usepackage{array}
\usepackage{multirow}
\usepackage{longtable}
\usepackage{framed} 
\usepackage{empheq} 
\usepackage[normalem]{ulem} 
\usepackage{enumitem}

\usepackage{epstopdf}
\usepackage{amsmath, amsfonts, amssymb}
\usepackage{amsthm}
\usepackage{yhmath}
\usepackage{epsf}
\usepackage{amsfonts}
\usepackage{stmaryrd}
\usepackage{amssymb}
\usepackage{leftidx}
\usepackage{mathtools}
\usepackage{placeins}
\usepackage{booktabs}
\usepackage{enumitem}
\usepackage{caption}
\usepackage{tabu}
\usepackage{multirow}
\usepackage{stackengine}
\usepackage[utf8]{inputenc}
\usepackage[english]{babel}
\usepackage{bm}
\usepackage{array}
\usepackage{multirow}
\usepackage{graphicx} 
\usepackage[T1]{fontenc}
\usepackage[utf8]{inputenc}
\usepackage{amssymb}
\usepackage{multicol}
\usepackage{bm}
\usepackage{graphicx}
\usepackage[export]{adjustbox}
\usepackage[font=footnotesize]{caption}
\usepackage{float} 
\usepackage{placeins}
\usepackage{longtable}

\usepackage{amsmath,stmaryrd,footmisc}
\usepackage{esint}
\usepackage{placeins}
\usepackage{algorithm}
\usepackage{algpseudocode}
\usepackage{esvect}
\MakeRobust{\vv}

\usepackage{epstopdf}
\usepackage{natbib}
\usepackage{nohyperref}
\usepackage{graphicx}
 \usepackage{setspace}
\usepackage{color}
\usepackage{verbatim}
 \usepackage[usenames,dvipsnames]{pstricks}
 \usepackage{epsfig}
 \usepackage{pst-grad} 
 \usepackage{pst-plot} 

\usepackage{etoolbox}
\usepackage{mathtools}
\preto\subequations{\ifhmode\unskip\fi}

\usepackage{enumerate}
\usepackage{enumitem}
\usepackage{amsmath}
\usepackage{placeins}
\usepackage{siunitx}
\usepackage{cleveref}
\usepackage{empheq, esdiff}

\usepackage{physics}

\makeatletter
\@ifpackageloaded{esvect}{%
  \DeclareRobustCommand{\vect}[1]{\vv{#1}}%
}{%
  \RequirePackage{amsmath}
  \DeclareRobustCommand{\vect}[1]{\overrightarrow{#1}}%
}
\makeatother

\def\nten#1{\mathbf{#1}}
\def\mcf{\text{m}_{\text{cf}}}
\def\lch{\text{l}_{\text{ch}}}
\def\Dbe{\vect{\Delta \beta^\eps}}
\def\Dbecc{\vect{\Delta \beta^\eps_\texttt{cc}}}
\def\stsMC{\sigma_{\mathtt{ts}}^\mathtt{MC}}
\def\scsMC{\sigma_{\mathtt{cs}}^\mathtt{MC}}

\def\stsHB{\sigma_{\mathtt{ts}}^\mathtt{HB}}
\def\scsHB{\sigma_{\mathtt{cs}}^\mathtt{HB}}

\def\WtsMC{{W}^\mathtt{MC}_{\mathtt{ts}}}
\def\WtsHB{{W}^\mathtt{HB}_{\mathtt{ts}}}
\def\WcsMC{{W}^\mathtt{MC}_{\mathtt{cs}}}
\def\WcsHB{{W}^\mathtt{HB}_{\mathtt{cs}}}

\def\sigmTwo{\sigma_{\mathtt{m,2}}}
\def\smTwo{s_{\mathtt{m,2}}}
\def\sigmax{\sigma_{\mathtt{max}}}
\def\sigmin{\sigma_{\mathtt{min}}}
\def\sigint{\sigma_{\mathtt{int}}}
\def\toct{\tau_{\mathtt{oct}}}
\def\smax{s_{\mathtt{max}}}
\def\smin{s_{\mathtt{min}}}
\def\sint{s_{\mathtt{int}}}
\def\tauoct{\tau_{\text{oct}}}

\def\bfu{{\bf u}}

\def\bfE{{\bf E}}

\def\bfI{{\bf I}}

\def\bfN{{\bf N}}

\def\bfX{{\bf X}}

\def\eps{\varepsilon}

\def\e0{\varepsilon_0}

\def\s0{\sigma_0}

\def\sts{\sigma_{\mathtt{ts}}}
\def\scs{\sigma_{\mathtt{cs}}}

\def\de{\delta^\varepsilon}
\def\ce{c_{\mathtt{e}}}

\DeclareMathAlphabet{\mathsfit}{T1}{\sfdefault}{\mddefault}{\sldefault}
\SetMathAlphabet{\mathsfit}{bold}{T1}{\sfdefault}{\bfdefault}{\sldefault}

\theoremstyle{plain}
\newtheorem{theorem}{Theorem}

\newtheorem{remark}[theorem]{Remark} 

\long\def\symbolfootnote[#1]#2{\begingroup%
\def\thefootnote{\fnsymbol{footnote}}\footnote[#1]{#2}\endgroup}

\begin{document}
\begin{frontmatter}

\title{The phase-field model of fracture incorporating Mohr-Coulomb, Mogi-Coulomb, and Hoek-Brown strength surfaces \vspace{0.1cm}}

\vspace{-0.1cm}

\author[purdue]{S. Chockalingam}
\ead{csenthil@purdue.edu}

\author[columbia]{Adrian Buganza Tepole}
\ead{ab6035@columbia.edu}

\author[gatech]{Aditya Kumar\corref{cor1}}
\ead{aditya.kumar@ce.gatech.edu}

\address[purdue]{Department of Mechanical Engineering, Purdue University, West Lafayette, IN 47907, USA \vspace{0.05cm}}

\address[columbia]{Department of Mechanical Engineering, Columbia University, New York, NY 10027, USA \vspace{0.05cm}}

\address[gatech]{School of Civil and Environmental Engineering, Georgia Institute of Technology, Atlanta, GA 30332, USA \vspace{0.05cm}}

\cortext[cor1]{Corresponding author}

\begin{abstract}

\vspace{-0.1cm}

Classical phase-field theories of brittle fracture capture toughness-controlled crack growth but do not account for the material's strength surface, which governs fracture nucleation in the absence of cracks. The phase-field formulation of Kumar et al. (2020) proposed a blueprint for incorporating the strength surface while preserving toughness-controlled propagation by introducing a nucleation driving force and presented results for the Drucker-Prager surface. Following this blueprint, Chockalingam (2025) recently derived a general driving-force expression that incorporates arbitrary strength surfaces. The present work implements this driving force within a finite-element framework and incorporates representative strength surfaces that span diverse mathematical and physical characteristics---the Mohr-Coulomb, 3D Hoek-Brown, and Mogi-Coulomb surfaces. Through simulations of canonical fracture problems, the formulation is comprehensively validated across fracture regimes, capturing (i) nucleation under uniform stress, (ii) crack growth from large pre-existing flaws, and (iii) fracture governed jointly by strength and toughness. While the strength surfaces examined here already encompass a broad range of brittle materials, the results demonstrate the generality and robustness of the proposed driving-force construction for materials governed by arbitrary strength surfaces.

\endkeyword

\end{abstract}

\end{frontmatter}

\section{Introduction}

A complete description of fracture in elastic brittle materials requires three ingredients: elastic function, toughness, and strength \cite{KBFLP20}. A mathematical theory of fracture that combines these has remained elusive. In its absence, various modeling approaches have been proposed over the past 60 years, but none have robustly integrated the three ingredients. The phase-field approach to fracture \cite{Francfort98, Bourdin00}---initially developed to address when and where cracks grow---has more recently been proposed as a framework capable of unifying the three necessary ingredients for a complete model. Within this framework, several models have been introduced, but the model by Kumar et al. \cite{KFLP18, KBFLP20} has proved most successful so far in describing fracture nucleation and propagation across a wide range of materials (including soft materials), loading conditions, and geometries \cite{KLP21, KRLP22, LK24, KKLP24, KDK2025Comparison}.\\

The phase-field approach of Kumar et al. (2018) \cite{KFLP18} was motivated by their observation that classical phase-field models could account for, at best, the tensile strength of a material. However, as laid out in \cite{KBFLP20}, the notion of strength extends beyond a single tensile value---it defines a surface in stress space that represents the set of critical stresses under arbitrary uniform multiaxial loadings in a large homogeneous specimen. This surface, termed the strength surface, had been missing from other phase-field formulations. Kumar et al. proposed that the strength surface could be incorporated into classical phase-field models by introducing an additional driving force into the partial differential equation governing the evolution of the phase-field variable. When suitably constructed, this driving force accurately captures the strength surface while leaving largely intact the ability of classical phase-field regularization to model crack propagation according to Griffith's fracture postulate \cite{KDK2025Comparison}. In fact, it improves upon the classical models in capturing crack growth under global compressive strains. The strength criterion acts as an effective constraint that prevents crack growth in compressive regions, as first discussed by Liu and Kumar \cite{LK24}.\\

The recipe for constructing the driving force corresponding to an arbitrary strength surface was first presented in Kumar et al. \cite{KBFLP20, KRLP22}; however, explicit formulas were provided only for the Drucker-Prager surface. The Drucker-Prager surface is defined by a linear function of the first two principal invariants of the stress tensor and can fit strength data well for many ceramics, such as graphite and titania \cite{KBFLP20}. However, other materials require different strength surfaces, such as Mohr-Coulomb,  Hoek-Brown, and Mogi-Coulomb for rocks and geomaterials \cite{labuz2014mohr,hoek1992modified,al2005relation}, and generalized Podgórski-Bigoni-Piccolroaz for elastomers \cite{rosendahl2019equivalent}. Chockalingam (2025) \cite{chockalingam2025construction} presented an explicit expression for the driving force for an arbitrary strength surface. The present study makes use of this expression to incorporate the following three representative and widely used strength surfaces: \textit{Mohr-Coulomb (M-C)}, \textit{3D Hoek-Brown (3D H-B)}, and \textit{Mogi-Coulomb (Mg-C)}. These surfaces collectively span a wide range of mathematical and physical characteristics, enabling a rigorous assessment of the generality and robustness of the driving-force formulation.\\

Historically, the Mohr-Coulomb (M-C) criterion \cite{labuz2014mohr} is one of the earliest and most widely adopted descriptions of strength failure \cite{singh2011modified}. It assumes a linear relation between normal and shear stress on the failure plane. Owing to its simplicity and clear physical interpretation, the M-C surface has been successfully applied to a broad range of materials including soils \cite{consoli2014mohr}, rocks \cite{chang2018application}, ice \cite{arenson2005mathematical,mamot2018iceMC,clayton2025MC}, mortar \cite{niwa1967mortarMC,ghimire2022determination}, concrete \cite{lelovic2020determination}, and bone \cite{wang2008identification}. The ability to model failure in these materials has direct practical implications---for instance, understanding ice-shelf breakup to study climate change, assessing the structural integrity of cementitious materials, and predicting bone fracture in biomedical contexts. Despite its wide successful use the M-C surface suffers from two maladies: (i) it does not capture the nonlinear relationship between the normal and the shear stress at larger stresses, and (ii) it only depends on the maximum and minimum principal stresses and ignores the influence of the intermediate principal stress which can significantly affect failure under polyaxial loadings in geomaterials \cite{murrell1963criterion,handin1967effects,hoskins1969failure,mogi1971fracture,takahashi1989effect,haimson2000new,haimson2002true}. The Hoek-Brown criterion was specifically developed to model rock failure and addresses the first of the limitations by accounting for the nonlinear relationship between normal and shear stress. However, it also ignores the intermediate principal stress. A `3D' extension of the Hoek-Brown criterion was proposed by Zhang and Zhu \cite{zhang2007three,zhang2008generalized}, which additionally incorporates the effect of the intermediate principal stress, providing a more accurate description of rock strength under true triaxial conditions.\\

The Mogi-Coulomb (Mg-C) criterion \cite{al2005relation,al2006stability} provides an alternative linear description (homogeneous function of stress of degree one) that includes the intermediate principal stress but reduces to the Mohr-Coulomb criterion in loadings that do not have an intermediate principal stress distinct from the maximum or minimum principal stress. Thus, it combines the simplicity of the M-C surface (and associated inability to capture nonlinear normal-shear relationship) with improved experimental fidelity. Together, the M-C, 3D H-B, and Mg-C surfaces represent some of the most widely employed strength surfaces in studying rock failure and thus  
are central to practical applications such as tunneling, borehole stability, underground excavation, and landslide prediction---each of which has major economic and safety significance. \\

While the M-C and Mg-C criteria are described by linear  stress functions (homogeneous function of degree one), the 3D Hoek-Brown (3D H-B) criterion requires a nonlinear stress function (non-homogeneous function). The combination of M-C, 3D H-B, and Mg-C surfaces thus provides an ideal set of test cases: they span linear and nonlinear stress functions and exhibit differing sensitivities to the intermediate principal stress. Thus, in addition to their applicability to a wide range of important brittle materials, their implementation allows us to assess both the generality and numerical robustness of the driving-force construction, providing confidence in the general applicability of the phase-field theory of Kumar et al.  (2020) \cite{KBFLP20} to a broad range of brittle materials with arbitrary strength surfaces. The purpose of this work is to present complete and comprehensively validated phase-field models of fracture that incorporate these surfaces. The models' ability to capture nucleation under uniform stress, crack growth from large pre-existing cracks, and the mediation between strength and toughness is demonstrated through comparison of finite element simulations with analytical results. We note that the fracture response of materials described by these strength criteria, such as rocks, is often quite complex, with several inelastic mechanisms that may be activated under compressive loading. In this work, however, we restrict our attention to an idealized elastic–brittle representation, as our objective is to validate the incorporation of strength criteria within the phase-field formulation rather than to provide a detailed constitutive description of specific materials. \\

The remainder of this paper is organized as follows. In \Cref{sec:Theory}, we introduce the strength-incorporated phase-field theory of Kumar et al. (2020) \cite{KBFLP20} and the driving force expression presented by Chockalingam (2025) \cite{chockalingam2025construction} for an arbitrary elastic brittle material. Subsequently, in \Cref{subsec:strength_surfaces}, we specialize the formulation for our specific strength surfaces of choice here. \Cref{Sec: Numerics} presents numerical results from finite element simulations that are validated against 
analytical results for various fracture problems. 
We provide some concluding remarks in \Cref{Sec: Final Comments}.

\section{Theory}\label{sec:Theory}

\subsection{Kinematics and basic definitions}
Consider a structure made of an isotropic linear elastic brittle material occupying an open bounded domain $\mathrm{\Omega}\subset \mathbb{R}^3$, with boundary $\partial\mathrm{\Omega}$ and outward unit normal $\nten{N}$, in its undeformed and stress-free configuration at time $t=0$. At a later time $t \in (0, T]$, due to an externally applied displacement $\overline{\bfu}(\bfX, t)$ on a part $\partial\mathrm{\Omega}_\mathcal{D}$ of the boundary and a traction $\overline{\textbf{t}}(\bfX,t)$ on the complementary part $\partial\mathrm{\Omega}_\mathcal{N}=\partial\mathrm{\Omega}\setminus \partial\mathrm{\Omega}_\mathcal{D}$, the material points in the structure described by position vector $\nten{X}$ experience a displacement described by the field $\nten{u}(\nten{X},t)$. We write the infinitesimal strain tensor as
\begin{equation*}
\bfE(\bfu)=\dfrac{1}{2}(\nabla\bfu+\nabla\bfu^T).
\end{equation*}
Material impenetrability enforces that ${\rm det}(\bfI+\nabla\bfu)>0$.  In response to the externally applied mechanical stimuli, cracks can also nucleate and propagate in the structure. Those are described in a regularized fashion via a phase field
\begin{equation*}
v=v(\bfX,t)
\end{equation*}
taking values in $[0,1]$. Precisely, $v=1$ identifies regions of the sound material, whereas $0\le v<1$ identifies regions of the material undergoing fracture.

\subsection{Constitutive behavior of the material}
\label{subsec:constit}
Based on decades of experimental observations, the mechanical behavior of a brittle material is assumed to be fully characterized by three intrinsic properties of the material: (\emph{i}) elasticity, (\emph{ii}) strength, and (\emph{iii}) toughness/critical energy release rate.\\

\noindent \textit{(i) Elasticity}: The elasticity for an isotropic linear elastic material is characterized by the strain-energy density function
\begin{equation}
	W(\bfE(\bfu)) =\mu \, {\rm tr}\,\bfE^2+\dfrac{\lambda}{2}({\rm tr}\,\bfE)^2,\label{W-mu}
\end{equation}
where $\mu>0$ and $\lambda>-2/3\mu$ are the Lam\'e constants. Recall the basic relations $\mu=E/(2(1+\nu))$ and $\lambda=E\nu/((1+\nu)(1-2\nu))$, where $E$ is the Young's modulus and $\nu$ is the Poisson's ratio. The Cauchy stress $\bm{\sigma}$ is given by
\begin{equation}
    \bm{\sigma} =  \pdv{W(\nten{E})}{\nten{E}} = \dfrac{E}{1+\nu}\bfE+\dfrac{E\,\nu}{(1+\nu)(1-2\nu)}({\rm tr}\,\bfE)\bfI\label{eq:Stress}
\end{equation}
The principal values of the Cauchy stress are taken to be given by $\sigma_1, \sigma_2, \sigma_3$ whereas the maximum and minimum values are denoted by $\sigma_\text{max} = \text{max}(\sigma_1, \sigma_2, \sigma_3)$ and $\sigma_\text{min} = \text{min}(\sigma_1, \sigma_2, \sigma_3)$. For subsequent use we define the following invariants of the stress
\begin{equation}
   I_1 = \sigma_1 + \sigma_2 + \sigma_3,\quad J_2 = \tfrac{1}{6}\left( (\sigma_1 - \sigma_2)^2 + (\sigma_2 - \sigma_3)^2 + (\sigma_3 - \sigma_1)^2 \right), \quad \sigma_{\mathtt{m,2}} = \frac{1}{2}\left(\sigmax + \sigmin\right)\label{eq:Invariants} 
\end{equation}
For convenience in describing rock mechanics, where the literature generally assumes compressive stress to be positive, we define the stress $\nten{s} = - \bm{\sigma}$ whose corresponding principal stresses are given by $s_1, s_2, s_3$ where $\smax = -\sigmin$ and $\smin = -\sigmax$. The intermediate principal stress $\sigint$ is given by $\sigint = I_1 - 2\sigma_{\mathtt{m,2}}$ and we can define $\sint=-\sigint$. We also note that the commonly used stress terms $\smTwo$  and $\toct$ (octahedral shear stress) in the literature on rock mechanics are given by $\smTwo = -\sigmTwo$ and $\toct = \sqrt{\frac{2}{3} J_2}$.\\ 

\noindent \textit{(ii) Strength surface: } The strength of the material controls fracture nucleation in large specimens of homogeneous brittle materials subjected to a uniform state of stress $\bm{\sigma}$. When the structure is subjected to a state of monotonically increasing and spatially uniform stress, a crack will form at an indeterminate location at a critical value of the applied stress. The set of all such critical stresses defines a surface in stress space. This surface is referred to as the strength surface of the material and is considered a material property (save for stochasticity) in the macroscopic theory here \citep{KBFLP20}. We assume that it can mathematically be described in the following general normalized form
\begin{equation}
    \mathcal{F} \equiv {F}(\bm{\sigma};\vect{\beta}) = g(\bm{\sigma};\vect{\beta}) - 1 = 0 \label{eq:SSurf-0}
\end{equation}
where $\vect{\beta}$ is an $n$-dimensional array of material constants/parameters. For any stress state $\bm{\sigma}$ before attainment of strength (not on the locus $\mathcal{F}$), we assume $\hat{F}({\bm{\sigma}}) < 0$ and that $\hat{F}({\bm{\sigma}}) > 0$ is in violation of the strength of the material.
Further, we assume that  $g(\bm{0};\vec{\beta})=0$ so that ${F}(\bm{0};\vect{\beta}) = -1$. Specific constitutive choices of the strength surface are discussed later in this section. For later use, we define the uniaxial tension and uniaxial compression strengths, denoted by $\sts$ and $\scs$, respectively, as the critical stress values on the strength surface when the material is loaded under uniaxial tension or compression.

{
\begin{remark}{\rm Some materials that are brittle in tension may show permanent deformation in compression before failure due to granular flow, pore collapse, crushing, fragmentation, and ductile shear banding \citep{anand2025fracture}, among other reasons. In that case, the surface (\ref{eq:SSurf-0}) for $I_1<0$ may represent the onset of the permanent deformation. Localization and increasing permanent deformation may follow, and crack formation may occur when a bounding surface in the stress-deformation space is reached.
}
\end{remark}
}

\noindent \textit{(iii) Fracture toughness}: The critical energy release rate (or fracture toughness), denoted as $G_c$, controls fracture nucleation from a large pre-existing crack, $\Gamma$, through the Griffith criticality condition
\begin{equation}
G = -\dfrac{\partial{\mathcal{W}}}{\partial \Gamma}=G_c \label{Griffith},
\end{equation}
where $\mathcal{W}$ is the potential energy of the structure and $G$ is the energy release rate---the reduction in potential energy with respect to an added surface area ${\partial \Gamma}$ to the crack. Thus, a large pre-existing crack may advance when the energy release rate reaches the fracture toughness $G_c$ of the material. Throughout this manuscript, the notion of a `large' crack refers to the crack (as well as body dimensions) being much larger than a characteristic fracture process zone size that represents the size of the locally strength-violated region near the crack tip. The Irwin characteristic length is such a measure of the fracture process zone size and is defined as follows for Mode I fracture 
\begin{equation}
    \lch = \frac{3 E G_c}{8 \sts^2} \label{eq:lch}
\end{equation}
For other fracture modes, the tensile strength $\sts$ in \cref{eq:lch} needs to be substituted with the appropriate strength.

\subsection{Governing equations of the strength-incorporated phase-field theory}

Under general loadings and non-uniform stress states, an interplay between strength and fracture toughness governs fracture behavior. The phase-field theory proposed by Kumar et al. \cite{KFLP18, KBFLP20} incorporates the strength of the material by way of introduction of a driving force $\ce(\nten{X},t)$ into the Euler-Lagrange equations of the classical phase-field theory \cite{Bourdin00,Bourdin08}, which already contain the Griffith criticality condition. %
The governing equations of the strength-incorporated phase-field theory are then written as follows: the displacement field $\bfu_k(\bfX)=\bfu(\bfX,t_k)$, and phase field $v_k(\bfX)=v(\bfX,t_k)$ at any material point $\bfX\in\overline{\mathrm{\Omega}} = \mathrm{\Omega} \cup \partial\mathrm{\Omega}$ and discrete time $t_k\in\{0=t_0,t_1,...,t_m,$ $t_{m+1},...,$ $t_M=T\}$ are determined by the following system of coupled partial differential equations:
\begin{subequations}\label{BVP-u-theory}
\begin{empheq}[left=\empheqlbrace]{align}
&{\rm Div}\!\left[v_{k}^2\dfrac{\partial W}{\partial \bfE}(\bfE(\bfu_{k}))\right]={\bf0}, \quad \text{if} ~ \bfX \in \Omega, \label{BVP-u-theory:a}\\[3pt]
&\bfu_{k}=\tilde{\bfu}(\bfX,t_{k}), \quad \text{if} ~ \bfX \in \partial\mathrm{\Omega}_\mathcal{D}\label{BVP-u-theory:b}\\[3pt]
&\left[v_{k}^2\dfrac{\partial W}{\partial \bfE}(\bfE(\bfu_{k}))\right]\bfN=\tilde{\textbf{t}}(\bfX,t_{k}), \quad \text{if} ~ \bfX \in \partial\mathrm{\Omega}_\mathcal{N} \label{BVP-u-theory:c}
\end{empheq}
\end{subequations}
and\\
\begin{subequations}\label{BVP-v-theory-ce}
\begin{empheq}[left=\empheqlbrace]{align}
&\dfrac{3}{4}\varepsilon \, \de \, G_c \triangle v_{k}
   = 2 v_{k} W(\bfE(\bfu_{k})) - \ce(\nten{X},t_k)
   - \dfrac{3}{8} \dfrac{ \de G_c}{\varepsilon},\quad
   \text{if } v_{k}(\bfX)< v_{k-1}(\bfX),\quad 
   \bfX\in\Omega, \label{BVP-v-theory_main}\\[4pt]
&\dfrac{3}{4}\varepsilon \, \de \, G_c \triangle v_{k}
   \ge 2 v_{k} W(\bfE(\bfu_{k})) - \ce(\nten{X},t_k)
   - \dfrac{3}{8} \dfrac{ \de G_c}{\varepsilon},~ 
   \text{if } v_{k}(\bfX)=1 \text{ or } v_{k}(\bfX)= v_{k-1}(\bfX)>0, 
   \bfX\in\Omega, \label{BVP-v-theory_secondary}\\[4pt]
&v_{k}(\bfX)=0,\quad 
   \text{if } v_{k-1}(\bfX)=0,\quad 
   \bfX\in\Omega, \label{BVP-v-theory_zero}\\[4pt]
&\nabla v_{k}\cdot\bfN=0,\quad 
   \bfX\in \partial\Omega. \label{BVP-v-theory_bc}
\end{empheq}
\end{subequations}
where $\triangle(\cdot) \equiv \text{Div}(\nabla(\cdot))$ and $\varepsilon>0$ is the regularization length---the length scale over which a sharp crack is smeared by the phase field. 
In practice, $\eps$ should be chosen so that it is not much larger than the fracture process zone size. 
$\de>0$ is a scalar correction factor that helps preserve correct fracture toughness in Griffith physics, as explained later in this section. 
In the phase-field equations above, the true stress $\bm{\sigma}$ is given by

\begin{equation}
    \bm{\sigma} = v^2 \pdv{W(\nten{E})}{\nten{E}} = v^2\left(\dfrac{E}{1+\nu}\bfE+\dfrac{E\,\nu}{(1+\nu)(1-2\nu)}({\rm tr}\,\bfE)\bfI\right) \label{eq:true_Stress}
\end{equation}

\subsubsection{Driving force $\ce$}
A roadmap for the construction of the driving force $\ce$ that incorporates the material strength was presented in Kumar et al. (2020) \cite{KBFLP20}, where explicit expressions were provided for the Drucker-Prager strength surface. Following this roadmap, Chockalingam (2025) \cite{chockalingam2025construction} derived the following expression for $\ce$ for a general strength surface described by \cref{eq:SSurf-0},
\begin{equation}
    \ce = -\omega_\eps g(\bm\sigma; \vect{\beta} + \Dbe) \quad \text{where} \quad \omega_\eps = \dfrac{3}{8}  \dfrac{\de \, G_c}{\varepsilon}, \label{eq:general_ce_sol}
\end{equation}
and $\bm\sigma$ is the degraded true stress (\ref{eq:true_Stress}). This driving force $\ce$ allows for the strength function $F$ of the material to be recovered exactly in the limit of vanishing regularization length $\eps \searrow 0$ while the correction parameters $\vv{\Delta \beta^\eps}$ allow for $n$ chosen strength states on the strength surface to be recovered even for any finite regularization length. Specifically, for chosen independent\footnote{See \cite{chockalingam2025construction} for a formal notion of what it means for the chosen strength calibration locations to be `independent'.} calibration strength states ${\bm{\sigma}}_{\texttt{s}_i}$ ($i =1,2,3,... , n$) on the material's strength surface, 
 the correction parameters $\vect{\Delta \beta^\eps}$ are obtained by solving the following set of coupled equations,
\begin{subequations}\label{eq:surfacematching}
\begin{empheq}[left=\empheqlbrace, right={\quad \text{for } i=1,2,3,\ldots,n}]{align}
& 2 \dfrac{W({\bm{\sigma}}_{\texttt{s}_i})}{\omega_\eps}
  + F({\bm{\sigma}}_{\texttt{s}_i}; \vect{\beta} + \vect{\Delta \beta^\eps}) = 0, \label{eq:surfacematching_a}\\[4pt]
& F({\bm{\sigma}}_{\texttt{s}_i}; \vect{\beta}) = 0. \label{eq:surfacematching_b}
\end{empheq}
\end{subequations}
where $W(\bm{\sigma})$ is the complementary strain energy function
\begin{equation}
    {W}(\bm{\sigma}) = \frac{1}{2} \left(\frac{{J}_2}{{\mu}} + \frac{{I}_1^2}{9 {K}}\right) \label{eq:straineenergy2}
\end{equation}
evaluated at $v=1$ and $K= \lambda + \frac{2\mu}{3}$ is the bulk modulus. When the strength function $F$ is linear in the parameters $\vect{\beta}$, the following explicit solution for $\Dbe$ can be derived \cite{chockalingam2025construction} 
\begin{equation}
  \vect{\Delta \beta^\eps} = -\left(\pdv{\vv{F}}{\vv{\beta}}\right)^{-1}\frac{2\vect{{W}}}{{\omega}_\eps} \quad \text{where} \quad  {W}_i =  {W}({\bm{\sigma}}_{\texttt{s}_i}), \quad \pdv{F_i}{{\beta}_j} = \pdv{F({\bm{\sigma}}_{\texttt{s}_i}; \vv{\beta})}{\beta_j} \label{eq:deltabetaeps_sol}
\end{equation}
 In the sharp limit ($\eps \searrow 0$), the parameters $\vv{\Delta \beta^\eps}$ vanish since $\omega_\eps \nearrow \infty$.\\

In the presence of large cracks (toughness-dominated regime), the system of equations (\ref{BVP-u-theory})-(\ref{BVP-v-theory-ce}) with the driving force $\ce$ defined above shows behavior consistent with Griffith's criticality condition (\ref{Griffith}); however, with a different effective critical energy release rate. The value of the effective critical energy release rate can be corrected to match the experimental value, $G_c$, through the scalar parameter $\de$. The parameter $\de$ can be numerically calibrated for any boundary-value problem of choice for which the nucleation from a large pre-existing crack can be determined exactly according to Griffith's equation (\ref{Griffith}).  Based on the numerical results presented in various studies using the Drucker-Prager strength surface \cite{KBFLP20, KRLP22, KLDLP24, KKLP24, LK24,KAMAREI2026118449}, $\de$ is observed to be independent of the boundary value problem under investigation. Later in the manuscript, we will further verify this observation for the different strength surfaces considered here. Further, an approximate analytical expression for $\de$ was provided in \cite{KKLP24} for the Drucker-Prager strength surface through a large number of numerical simulations. We will later provide such formulas for $\de$ for the strength surfaces considered here.

\begin{remark}
   \rm Note that setting $\ce =0$ and $\de=1$ in the governing equations (\ref{BVP-u-theory})-(\ref{BVP-v-theory-ce}) recovers the classical phase-field theory that captures Griffith physics of large cracks but does not encode the material strength surface. Hence, the computational implementation of the equations (\ref{BVP-u-theory})--(\ref{BVP-v-theory-ce}) only differs from the implementation of the classical variational model through the presence of those additional terms.
\end{remark} 

\begin{remark} \rm
Substituting the expression for the driving force $\ce$ (\cref{eq:general_ce_sol}) into the governing equations (\ref{BVP-v-theory-ce}) for the phase field, we can show that the relevant equations \cref{BVP-v-theory_main,BVP-v-theory_secondary} can be written in a simplified manner resembling the classical phase-field theory as follows \cite{chockalingam2025construction}
\begin{subequations}\label{BVP-v-theory_EGc}
\begin{empheq}[left=\empheqlbrace]{align}
&\dfrac{3}{4}\varepsilon\,\de\,G_c\,\triangle v_{k}
   = 2v_{k}W(\bfE(\bfu_{k})) 
   - \dfrac{3}{8}\dfrac{\hat{G}^\varepsilon_c}{\varepsilon},\quad
   \mbox{if } v_{k}(\bfX)<v_{k-1}(\bfX),\quad 
   \bfX\in\Omega, \label{BVP-v-theory_main_EGc}\\[4pt]
&\dfrac{3}{4}\varepsilon\,\de\,G_c\,\triangle v_{k}
   \ge 2v_{k}W(\bfE(\bfu_{k}))
   - \dfrac{3}{8}\dfrac{\hat{G}^\varepsilon_c}{\varepsilon},\quad  
   \mbox{if } v_{k}(\bfX)=1\;\mbox{or}\;
   v_{k}(\bfX)=v_{k-1}(\bfX)>0,\quad 
   \bfX\in\Omega. \label{BVP-v-theory_secondary_EGc}
\end{empheq}
\end{subequations}
where the effective fracture toughness $\hat{G}_c^\eps$ is given by
\begin{equation}
    \hat{G}_c^\eps = -\de F({\bm{\sigma}}; \vv{\beta} + \vv{\Delta \beta^\eps}) G_c
\end{equation}
Setting $\hat{G}_c^\eps=G_c$ and $\de=1$ recovers the governing equations of the classical phase-field theory.
\end{remark}

\noindent \textit{The phase-field strength surface:} The strength surface ${\mathcal{F}}^\eps$ generated by the strength-incorporated phase-field theory is given by the following equation  \citep{KBFLP20,chockalingam2025construction},
\begin{equation}
 {\mathcal{F}}^\eps \equiv   \frac{2 W({\bm{\sigma}})}{\omega_\eps} + {F}({\bm{\sigma}}; \vect{\beta} + \vect{\Delta \beta^\eps}) = 0 \label{eq:First_phase_ss_general}
\end{equation}
 Note that setting $v_k=1$ in \cref{BVP-v-theory_main} yields \cref{eq:First_phase_ss_general}. Numerical results \cite{KBFLP20, KLP20} have indicated that for sufficiently large structures, the initially uniform phase field solution loses stability and localizes into a crack when the condition above is satisfied for arbitrary multiaxial displacement-controlled loadings. We shall henceforth refer to \cref{eq:First_phase_ss_general} as the phase-field strength surface. Note that by construction, the phase-field strength surface approaches the material's strength surface ($\mathcal{F}^\eps \to \mathcal{F}$) in the limit of vanishing regularization length $\varepsilon \searrow 0$ since $\omega_\eps \nearrow \infty$ and $\Dbe \searrow 0$.\\

\noindent \textit{Correction in compression regime:} While the phase-field strength surface exactly recovers the material strength surface in the limit of vanishing regularization length, the mismatch for finite values of regularization lengths used in simulations can become quite large in the compressive regions ($I_1<0$), as we will demonstrate later. Consequently, a remedy was proposed in \cite{KRLP22} that adds a correction term to the driving force to improve the strength surface representation for $I_1<0$. This correction modifies the governing equations \cref{BVP-v-theory_main_EGc,BVP-v-theory_secondary_EGc} as follows
\begin{subequations}\label{BVP-v-theory_I1corrected}
\begin{empheq}[left=\empheqlbrace]{align}
&\dfrac{3}{4}\varepsilon\,\de\,G_c\,\triangle v_{k}
   = 2v_{k}W(\bfE(\bfu_{k}))\,f_\texttt{cc}(I_1)
   - \dfrac{3}{8}\dfrac{\hat{G}^\varepsilon_c}{\varepsilon},\quad
   \mbox{if } v_{k}(\bfX)<v_{k-1}(\bfX),\quad 
   \bfX\in\Omega, \label{BVP-v-theory_I1corrected1}\\[4pt]
&\dfrac{3}{4}\varepsilon\,\de\,G_c\,\triangle v_{k}
   \ge 2v_{k}W(\bfE(\bfu_{k}))\,f_\texttt{cc}(I_1)
   - \dfrac{3}{8}\dfrac{\hat{G}^\varepsilon_c}{\varepsilon},\quad  
   \mbox{if } v_{k}(\bfX)=1\;\mbox{or}\;
   v_{k}(\bfX)=v_{k-1}(\bfX)>0,\quad 
   \bfX\in\Omega. \label{BVP-v-theory_I1corrected2}
\end{empheq}
\end{subequations}
where
\begin{equation}
    \quad f_\texttt{cc}({I}_1) = \begin{cases}
      1 & {I}_1 \ge 0 \\
      0 & {I}_1 < 0
  \end{cases}
\end{equation}
The parameters $\vect{\Delta \beta^\eps}$ and the phase-field strength surface are now given by the following equations in the compression-corrected form \cite{chockalingam2025construction}
\begin{equation}
  \vect{\Delta \beta^\eps_{\mathtt{cc}}} = -\left(\pdv{\vv{F}}{\vv{\beta}}\right)^{-1}\frac{2\vect{W'}}{{\omega}_\eps} \quad \text{where} \quad  {W}_i' = f_\texttt{cc}({I}_1({\bm{\sigma}}_{\texttt{s}_i})) {W}_i, \label{eq:deltabetaeps_sol_I1corrected}
\end{equation}
\begin{equation}
 {\mathcal{F}}^\eps_{\mathtt{cc}} \equiv   \frac{2 W({\bm{\sigma}})}{\omega_\eps} f_\texttt{cc}(I_1) + {F}({\bm{\sigma}}; \vect{\beta} + \vect{\Delta \beta^\eps_\mathtt{cc}}) = 0, \label{eq:phase_ss_general_I1}
\end{equation}
while all other equations remain unchanged. Thus, it is seen that the strain energy term driving the fracture is switched off in the compressive regime, where it no longer disturbs the phase-field strength surface.

\vspace{0.3cm}
\begin{remark}
   \rm In the strength-incorporated phase-field equations (\ref{BVP-u-theory})-(\ref{BVP-v-theory-ce}), the tension-compression asymmetry in crack growth arises naturally from the asymmetry of the strength surface. Because the compressive strength of brittle materials is typically much higher than their tensile strength, this asymmetry alone is sufficient to prevent compressive cracking, as discussed by Liu and Kumar \cite{LK24}. However, the correction in the compression regime helps obtain the asymmetry for finite regularization lengths.
\end{remark} 

\section{Choice of strength surfaces}
\label{subsec:strength_surfaces}

We now write the functional forms of our specific choices for the strength surface and derive the associated driving forces. We begin with the Mohr-Coulomb surface in \Cref{subsec:M-C_criterion}, whose stress function is linear (homogeneous of degree one) and ignores the intermediate principal stress. Subsequently, in \Cref{subsec:3DHB_surface} we introduce the Hoek-Brown surface and its 3D variant whose stress functions are nonlinear (non-homogeneous in stress). The Hoek-Brown criterion ignores the intermediate principal stress, whereas the 3D Hoek-Brown does not. Finally, we discuss the linear Mogi-Coulomb strength criterion in \Cref{subsec:Mogi_surface} that employs all three principal stresses but reduces to the Mohr-Coulomb criterion in loadings that do not have a distinct intermediate principal stress.

\subsection{Mohr-Coulomb} \label{subsec:M-C_criterion}

\begin{figure}[h]
    \centering   \includegraphics[width=0.5\textwidth]{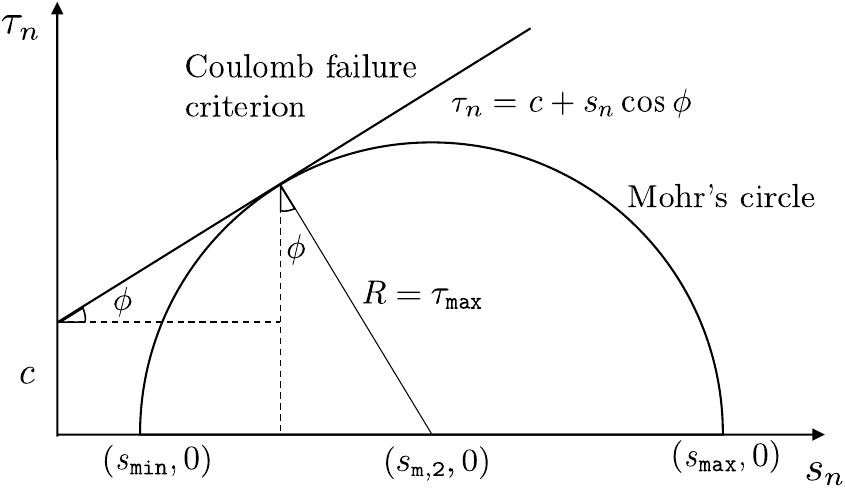}
    \caption{Illustration of the Mohr-Coulomb criterion. Failure occurs when the shear stress on any plane exceeds the sum of the cohesion $c$ of the material and the frictional resistance arising from the normal compression stress clamping the plane.}
    \label{fig:MC-illustration}
\end{figure}

In 1773, Coulomb proposed a failure criterion which suggested that a material will first fail when the shear stress $\tau_n$ acting on any internal plane overcomes the sum of (i) the material's cohesion $c$ and (ii) the frictional resistance generated by the compressive normal stress $s_n$ clamping that plane. Accordingly 
\begin{equation}
    \tau_n = c + \mu_f s_n, \quad \mu_f = \tan \phi \label{eq:MC_tauphi}
\end{equation}
where $\mu_f$ is the coefficient of friction defined in terms of the angle of internal friction $\phi$. The first attainment of this failure criterion can be obtained by solving for when the failure envelope defined by \cref{eq:MC_tauphi}, a straight line in the $s_n-\tau_n$ space at an angle $\phi$ to the $s_n$ axis, becomes tangential to the largest Mohr's circle defined by the pair of compressive principal stresses $\{\smax,\smin\}$ (see \Cref{fig:MC-illustration}). The center of the Mohr's circle is given by the coordinates $(s_n,\tau_n) = (\smTwo,0)$ where $\smTwo = \frac{1}{2}\left(\smax+\smin\right)$ and the radius of the Mohr's circle is given by $R=\tau_\mathtt{max}= \frac{1}{2}\left(\smax-\smin\right)$, where $\tau_\mathtt{max}$ is the maximum shear stress. The coordinates of the point where the failure envelope \cref{eq:MC_tauphi} becomes tangential to the Mohr's circle can then be written as,
\begin{equation}
\left(s_n,\tau_n\right) = \left(\smTwo - \tau_\mathtt{max} \sin \phi, \tau_\mathtt{max} \cos \phi \right),
\end{equation}
which, when plugged in \cref{eq:MC_tauphi} yields the following form of the Mohr-Coulomb (M-C) failure criterion 
\begin{equation}
 \frac{1}{2}\,(s_{\max} - s_{\min})
=
c\,\cos\phi
+
\frac{1}{2}\,(s_{\max} + s_{\min})\,\sin\phi%
\end{equation}
This equation can be rewritten in the following form
\begin{equation}
    \smax = \frac{2 c \cos \phi}{1-\sin \phi} + \left(\frac{1+\sin\phi}{1-\sin\phi}\right)\smin \label{eq:MC_sform1}
\end{equation}
The uniaxial tensile strength $\stsMC$ can be obtained by setting $\smin=-\stsMC$ and $\smax=0$ while the uniaxial compressive strength can be obtained by setting $\smax=\scsMC$, $\smin = 0$, yielding the equations
\begin{equation}
\stsMC = \frac{2c\cos\phi}{1+\sin\phi}=\frac{2c}{\tan{\left(\pi/4+\frac{\phi}{2}\right)}},\quad  \scsMC = \frac{2c\cos\phi}{1-\sin\phi}={2c}~{\tan{\left(\pi/4+\frac{\phi}{2}\right)}} 
\end{equation}
Thus, the M-C criterion in \cref{eq:MC_sform1} can be written in terms of principal components of compressive stress and the uniaxial tensile and compressive strengths as 
\begin{equation}
    \smax = \scsMC + \frac{\scsMC}{\stsMC}\smin \label{eq:MC_sform2}
\end{equation}
Written in terms of the conventional stress $\bm{\sigma}$ (positive in tension) and $\vect{\beta} = [\beta_1,\beta_2]$, the M-C criterion reads
    \begin{equation}
 \boxed{  \mathcal{F}_{\mathtt{MC} }\equiv {F}_{\texttt{MC}}({\bm{\sigma}}; \vect{\beta})  = \beta_1 \sigmax + \beta_2 \sigmin - 1 = 0, \quad \beta_1 = \frac{1}{\stsMC}, \ \beta_2 = -\frac{1}{\scsMC} }\label{eq:generalform_MC} 
\end{equation}
Thus, the intermediate principal stress does not play a role in the M-C strength criterion.\\

\begin{figure}[h]
    \centering   \includegraphics[width=0.45\textwidth]{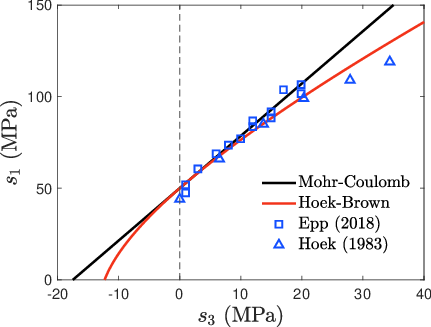}
    \caption{Experimental strength data for Indiana Limestone from conventional triaxial compression testing and fitted Mohr-Coulomb (\cref{eq:MC_s1s3}) and Hoek-Brown (\cref{eq:HBTC_In}) strength criteria. The axial compression stress $s_1$ is plotted against the confining pressure $s_3$.}
    \label{fig:LimestoneStrength}
\end{figure}

As described earlier in the Introduction, the M-C surface is often used to model the experimental strength data of rocks. The most commonly conducted strength test in rocks is the conventional triaxial compression test where in addition to a controlled confining pressure $p$, an increasing axial compression load is applied on the material until failure. For such a loading, identifying the axial compressive stress as $s_1$, we have $s_1 > s_2 = s_3 = p$. Hence $\smax = s_1$ and $\smin=s_3$. The M-C equation in \cref{eq:MC_sform1}  can then be written for the conventional triaxial compression test as
\begin{equation}
    s_1 = \scsMC + \frac{\scsMC}{\stsMC}s_3 \label{eq:MC_s1s3}
\end{equation}
When plotted in the plane of $s_1-s_3$, this defines a straight line with slope equal to the ratio of uniaxial compressive to tensile strength, vertical intercept equal to the uniaxial compressive strength, and horizontal intercept equal to the negative of the uniaxial tensile strength. Experimental strength data from conventional triaxial compression tests on several rocks often show such a linear relationship, particularly in the regime of low confining pressures. In \Cref{fig:LimestoneStrength}, we plot the experimental strength data for Indiana Limestone from triaxial tests done by Epp (2018) \cite{epp2018laboratory} and by Hoek (1983) \cite{hoek1983strength} (datapoints extracted from the report by Li et al. (2024) \cite{li2024stress}) up to higher confining pressures. Indiana Limestone has been selected as a representative material for all our analysis in this manuscript. It is a common benchmark material in rock mechanics, is widely quarried, extensively used as a building material in monumental public structures, and has been mechanically tested exhaustively.\\

The Mohr-Coulomb criterion in \cref{eq:MC_s1s3} models the experimental strength behavior of Limestone quite well in the low confining pressure regime where the $s_1-s_3$ relationship is nearly linear, as shown in \Cref{fig:LimestoneStrength}. The Mohr-Coulomb material parameters fitted in this regime are given by 
\begin{equation}
    \scsMC = 50\ \text{MPa},\ \stsMC = 17.5\ \text{MPa} \quad \equiv\quad c = 14.8 \text{MPa},\ \phi = 28.72\degree  
\end{equation}
However, it is observed that at higher confining pressures, the M-C criterion is inadequate. This is because the internal friction angle $\phi$ (or equivalently the coefficient of friction) and cohesion $c$ are no longer constant under high confining pressures. Additionally, the criterion ignores the influence of the intermediate principal stress. In tests where $\smax=\sint$ or $\sint=\smin$ (such as conventional triaxial compression tests), there is no distinct intermediate principal stress and thus its role in the strength behavior gets masked. However truly polyaxial strength tests with distinct intermediate principal stress have demonstrated that the intermediate principal stress does affect the strength criterion \cite{murrell1963criterion,handin1967effects,hoskins1969failure,mogi1971fracture,takahashi1989effect,haimson2000new,haimson2002true}.  Further, the criterion also typically overpredicts the uniaxial tensile strength when fitted to conventional triaxial compression test data. The strength criteria discussed next address these limitations of the M-C criterion. We note that despite these shortcomings, the M-C surface is one of the most widely used strength criteria in geotechnical engineering \cite{singh2011modified,labuz2014mohr} as well as in several other applications for various materials \cite{niwa1967mortarMC,arenson2005mathematical,clayton2025MC,wang2008identification,consoli2014mohr, mamot2018iceMC, ghimire2022determination,lelovic2020determination} due to its relative simplicity and underlying physical basis.\\

\noindent \textit{M-C surface incorporation into phase-field theory:} If the uniaxial tensile and compressive strengths are chosen as the  calibration locations on the strength surface (that are exactly captured for all $\eps$), the solution for the parameters $\Dbe$  can be calculated using \cref{eq:deltabetaeps_sol} (see \cite{chockalingam2025construction} for further details) as
\begin{equation}
    \Delta \beta^\eps_1 = -\frac{2 \WtsMC}{\stsMC\omega_\eps}, \quad \Delta \beta^\eps_2 = \frac{2 \WcsMC}{\scsMC\omega_\eps} \label{eq:Dbe_MC}
\end{equation}
where $\WtsMC$ and $\WcsMC$ are the values of the strain energy density function at the uniaxial tensile strength and uniaxial compressive strength states, which can be written using \cref{eq:straineenergy2} and the relation $E = \frac{9 \mu K}{3K+\mu}$ as
\begin{equation}
    \WtsMC = \frac{{\stsMC}^2}{2 E}, \quad \WcsMC = \frac{{\scsMC}^2}{2 E}
\end{equation}
The corresponding phase-field strength surface can be written using \cref{eq:First_phase_ss_general} as 
\begin{equation} 
 {\mathcal{F}}^\eps_\mathtt{MC} \equiv   \frac{2 W({\bm{\sigma}})}{\omega_\eps} + {F}_\mathtt{MC}({\bm{\sigma}}; \vect{\beta} + \vect{\Delta \beta^\eps}) = 0 \label{eq:PFSS_MC}
\end{equation}
where $\vect{\beta}$ and $\Dbe$ are given in \cref{eq:Dbe_MC,eq:generalform_MC}. For the compression corrected solution, the parameters $\Dbecc$  can be calculated using \cref{eq:deltabetaeps_sol_I1corrected}  as
\begin{equation}
    \Delta \beta^\eps_{\mathtt{cc}_1} = -\frac{2 \WtsMC}{\stsMC\omega_\eps}, \quad \Delta \beta^\eps_{\mathtt{cc}_2} = 0
\end{equation} 
and the phase-field strength surface from \eqref{eq:phase_ss_general_I1} as
\begin{equation}
 {\mathcal{F}}^\eps_\mathtt{MC, cc} \equiv   \frac{2 W({\bm{\sigma}})}{\omega_\eps}f_\texttt{cc}(I_1) + {F}_\mathtt{MC}({\bm{\sigma}}; \vect{\beta} + \vect{\Delta \beta_\mathtt{cc}^\eps}) = 0 \label{eq:PFSS_MC_cc}
\end{equation}
In \Cref{subsec:StrengthLimit}, plots of the phase-field strength surfaces defined above are compared with finite element simulation results of strength failure under multiaxial stress.\\

\noindent \textit{Expression for calibrated $\de$}: As mentioned earlier, the parameter $\de$ needs to be numerically calibrated from any boundary-value problem of choice for which the nucleation from a large pre-existing crack can be determined exactly according to Griffith's equation (\ref{Griffith}). This calibration needs to be performed for a given choice of material parameters and regularization length $\eps$. To facilitate use of the theory, a large number of calibration simulations were performed (using the single-edge notched tension test described later in \Cref{sec:SENT}) by varying material parameters and regularization lengths, and the following approximate functional form was fitted to the calibrated $\de$ values
\begin{equation}
    \de \approx \frac{0.3}{\mcf^2}\left(\frac{\lch^\texttt{MC}}{\eps}\right) + \frac{0.65}{\mcf} , \quad \lch^\texttt{MC} = \frac{3 E G_c}{8 {\stsMC}^2} = \frac{3G_c}{16\WtsMC} \label{eq:de_fitted_MC}
\end{equation}
where $\lch^\texttt{MC}$ is the Irwin characteristic length and $\mcf$ is the mesh correction factor that is a function of the characteristic mesh element size $h$ in the fracturing regions, given by,
\begin{equation}
    \mcf = 1+\frac{3 h}{8 \eps} \label{eq:mcf}
\end{equation}
The expression is valid in the range $0.5 \lesssim {\lch^\texttt{MC}}/{\eps} \lesssim 10$ and for both the regular and compression corrected versions of the theory.

\subsection{Hoek-Brown}\label{subsec:3DHB_surface}

Hoek and Brown developed the most influential strength criterion for rocks by empirical curve fitting of strength data from several hundred sets of rock triaxial test data and a large number of field tests \cite{hoek1980empirical,hoek1983strength,HoekBrown1988}. In its original form, the strength criterion is defined by the equation \cite{hoek1980empirical,HoekBrown1988}
\begin{equation}
    \smax = \smin + \sigma_c \left( m \frac{\smin}{\sigma_c} + s \right)^{\frac{1}{2}} \label{eq:HB_original}
\end{equation}
where $\sigma_c$ is the uniaxial compressive strength of the intact rock, and the parameters $m$ and $s$ are constants that depend on the properties of the rock and the extent of prior damage. For intact rock $m=m_i$ and $s=1$. This criterion was later modified \cite{hoek1992modified} to the current generalized form,
\begin{equation}
    \smax = \smin + \sigma_c \left( m_b \frac{\smin}{\sigma_c} + s \right)^{a} \label{eq:HB_gen}
\end{equation}
where $m_b$ is the value of the constant $m$ for a jointed rock mass and $s$ and $a$ are constants that depend on the characteristics of the rock mass. Empirical relations for the parameters $m_b$, $s$, and $a$ are provided in terms of the Geological Strength Index (GSI) as follows \cite{hoek2002hoek}:
\begin{equation}
m_b = \exp\left(\frac{\mathrm{GSI} - 100}{28 - 14D}\right) m_i,\quad s = \exp\left(\frac{\mathrm{GSI} - 100}{9 - 3D}\right),\quad a = 0.5 + \frac{1}{6} \left[ \exp\!\left(-\frac{\mathrm{GSI}}{15}\right) - \exp\!\left(-\frac{20}{3}\right) \right]
\end{equation}
where $D$ is the excavation disturbance factor that depends on the degree of disturbance due to blast damage and stress relaxation, and GSI reflects the structural surface characteristics of the rock mass. Note that for intact rock GSI $ = 100$ and thus $m_b=m_i$, $s=1$ and $a=0.5$. Due to its empirical nature and fitting to massive amounts of conventional triaxial testing, the criterion enjoys high prediction accuracy under typical stress loadings in rocks. Unlike in the Mohr-Coulomb criterion, the relationship between $\smax$ and $\smin$ is nonlinear in the H-B criterion. Equivalently, one can consider the cohesion $c$ and friction angle $\phi$ in \cref{eq:MC_tauphi} to now be stress dependent. \\

Once again, the uniaxial tensile strength $\stsHB$ can be obtained by setting $\smin=-\stsHB$ and $\smax=0$ while the uniaxial compressive strength can be obtained by setting $\smax=\scsHB$, $\smin = 0$, yielding the equations (following some algebra)
\begin{equation}
 \scsHB =  s^a \sigma_c ,\quad m_b s^{a-1} = \alpha_{\texttt{ct}} - \alpha_{\texttt{ct}}^{1-\frac{1}{a}} \quad \text{where} \quad \alpha_{\texttt{ct}} = \frac{\scsHB}{\stsHB} \label{eq:stsscs_HB}
\end{equation}
which specialize in intact rocks as
\begin{equation}
 \scsHB =  \sigma_c ,\quad m_i = \alpha_{\texttt{ct}} - \frac{1}{\alpha_{\texttt{ct}}}. \label{eq:stsscs_HB_intact}
\end{equation}
Recall from the previous section that for conventional triaxial compression testing, we have $\smax = s_1$  and $\smin = s_2 = s_3 = p$ where $s_1$ and $s_3$ are the axial compressive stress and confining pressure, respectively. Thus, the generalized Hoek-Brown criterion, when written for this test, becomes
\begin{equation}
     s_1 = s_3 + \sigma_c \left( m_b \frac{s_3}{\sigma_c} + s \right)^{a},
\end{equation}
which, when written for intact rock, specializes as
\begin{equation}
    s_1 = s_3 + \sigma_c \left( m_i \frac{s_3}{\sigma_c} + 1 \right)^{\frac{1}{2}}. \label{eq:HBTC_In}
\end{equation}
\\

The Hoek-Brown strength curve from \cref{eq:HBTC_In} fitted to the experimental strength data of intact Limestone introduced in the previous section is shown in \Cref{fig:LimestoneStrength}. The fitted parameter values are given by
\begin{equation}
  \stsHB = 12.3 \text{\ MPa}, ~ \scsHB = 50 \text{\ MPa} \quad \equiv \quad \sigma_c = 50 \text{\ MPa}, ~ m_i = 3.819 
\end{equation}
It is seen from \Cref{fig:LimestoneStrength} that the Hoek-Brown criterion is better than the Mohr-Coulomb criterion at capturing the experimental strength behavior of Limestone, particularly for higher confining pressures. This is because the Hoek-Brown criterion captures the nonlinear relationship between $s_1$ and $s_3$, or equivalently, the stress dependence of cohesion $c$ and friction angle $\phi$. It also predicts a lower uniaxial tensile strength, which is in line with the few reports of direct tensile strength measurements of the order of $5$ MPa \cite{schmidt1976fracture}. 
\\

\noindent \textit{3D Hoek-Brown criterion:} Despite the enormous success of the Hoek-Brown criterion in modeling rock behavior---as also shown above by its ability to capture the conventional triaxial compression strength data of Limestone---it shares a critical limitation with the Mohr-Coulomb criterion in that it neglects the influence of the intermediate principal stress, which, as discussed earlier, plays a significant role under truly multiaxial stress states \cite{murrell1963criterion,handin1967effects,hoskins1969failure,mogi1971fracture,takahashi1989effect,haimson2000new,haimson2002true}.  To address this limitation, Zhang and Zhu \cite{zhang2007three,zhang2008generalized} developed a `3D' version\footnote{The 3D refers to the dimensions of the principal stress space, the Zhang-Zhu criteria use all 3 principal stresses.} of the Hoek-Brown criterion such that it includes the intermediate principal stress, but reduces to the Hoek-Brown criterion in loading states where there is no distinct intermediate principal stress. The original form of the Zhang-Zhu criterion \cite{zhang2007three} extends the original form of the Hoek-Brown criterion in \cref{eq:HB_original} whereas the subsequent modified form \cite{zhang2008generalized} extends the generalized Hoek-Brown criterion in \cref{eq:HB_gen}. This generalized version, often referred to as the Generalized-Zhang-Zhu (GZZ) criterion in the literature, is written as  
\begin{equation}
    \frac{1}{{\sigma_c}^{\frac{1}{a} - 1}}
\left( \frac{3}{\sqrt{2}} \, \tauoct \right)^{1/a}
+ \frac{m_b}{2} \left( \frac{3}{\sqrt{2}} \, \tauoct \right) 
- m_b \smTwo
= s  \sigma_c \label{eq:GZZ}
\end{equation}
where the stress $\tauoct$ defined in \Cref{subsec:constit} depends on the intermediate principal stress and the parameters $m_b$ and $\sigma_c$ are the same as defined for the Hoek-Brown criterion. Note that the uniaxial tensile and compressive strength are the same as that for the Hoek-Brown surface, and the relations in \cref{eq:stsscs_HB,eq:stsscs_HB_intact} continue to hold. From here onwards, we will simply refer to GZZ as the 3D Hoek-Brown (3D H-B) criterion. \\

To summarize its advantages, the 3D Hoek-Brown criterion (i) reduces to the Hoek-Brown criterion for loading states where there is no distinct intermediate principal stress ($\smax=\sint$ or $\sint=\smin$), (ii) uses the same material parameters as the Hoek-Brown criterion (thus allowing direct use of extensive literature of calibrated material parameters) and (iii) offers improved match with strength data from experiments with a distinct intermediate principal stress \cite{zhang2007three,zhang2008generalized}. Therefore, we will directly make use of the 3D Hoek-Brown criterion in our analysis in this manuscript. The criterion can be rewritten as a linear function of two material parameters $\beta_1, \beta_2$ (with $\vect{\beta} = [\beta_1,\beta_2]$) as follows
\begin{equation}
\boxed{ \mathcal{F}_{\mathtt{HB} }\equiv {F}_{\texttt{HB}}({\bm{\sigma}}; \vect{\beta})  =    \beta_1 \left( \frac{3}{\sqrt{2}} \, {\tau_{\text{oct}}} \right)^{1/a}
+ \beta_2 \left( \frac{3}{2\sqrt{2}} \, \tau_{\text{oct}} - \smTwo\right) 
- 1 = 0} \label{eq:generalform_HB}
\end{equation} 
where
\begin{equation}
\beta_1 = \frac{1}{s\sigma_c^{\frac{1}{a}}} = \frac{1}{{\scsHB}^{\frac{1}{a}}}, \quad \beta_2 =\frac{m_b}{s \sigma_c} =\frac{1}{\stsHB}\left(1 - \left(\frac{\stsHB}{\scsHB}\right)^{\frac{1}{a}}\right) \label{eq:betas_HB}
\end{equation}\\

\noindent \textit{3D H-B surface incorporation into phase-field theory: } If the uniaxial tensile and compressive strengths are chosen as the  calibration locations on the strength surface, the solution for the parameters $\vect{\Delta \beta^\eps}$  can be calculated using \cref{eq:deltabetaeps_sol}  as
\begin{equation}
     \Delta \beta^\eps_1 = -\frac{2 \WcsHB}{{\scsHB}^\frac{1}{a}\omega_\eps}, \quad \Delta \beta^\eps_2 =\left(\frac{\stsHB}{\scsHB}\right)^{\frac{1}{a}} \frac{2 \WcsHB}{\stsHB\omega_\eps} -\frac{2 \WtsHB}{\stsHB\omega_\eps} \label{eq:Dbe_HB}
\end{equation}
where $\WtsHB$ and $\WcsHB$ are the values of the strain energy density function at the uniaxial tensile strength and uniaxial compressive strength states, which can be written using \cref{eq:straineenergy2} and the relation $E = \frac{9 \mu K}{3K+\mu}$ as
\begin{equation}
\WtsHB = \frac{{\stsHB}^2}{2 E}, \quad \WcsHB = \frac{{\scsHB}^2}{2 E}
\end{equation}
The corresponding phase-field strength surface can be written using \cref{eq:First_phase_ss_general} as 
\begin{equation}
 {\mathcal{F}}^\eps_\mathtt{HB} \equiv   \frac{2 W({\bm{\sigma}})}{\omega_\eps} + {F}_\mathtt{HB}({\bm{\sigma}}; \vect{\beta} + \vect{\Delta \beta^\eps}) = 0 \label{eq:PFSS_HB}
\end{equation}
where $\vect{\beta}$ and $\Dbe$ are given in \cref{eq:betas_HB,eq:Dbe_HB}. For the compression corrected solution, the parameters $\Dbecc$  can be calculated using \cref{eq:deltabetaeps_sol_I1corrected}  as
\begin{equation}
     \Delta \beta^\eps_{\mathtt{cc}_1} = 0, \quad \Delta \beta^\eps_{\mathtt{cc}_2} = -\frac{2 \WtsHB}{\stsHB\omega_\eps}
\end{equation} 
and the phase-field strength surface from \eqref{eq:phase_ss_general_I1} as
\begin{equation}
 {\mathcal{F}}^\eps_\mathtt{HB, cc} \equiv   \frac{2 W({\bm{\sigma}})}{\omega_\eps}f_\texttt{cc}(I_1) + {F}_\mathtt{HB}({\bm{\sigma}}; \vect{\beta} + \vect{\Delta \beta_\mathtt{cc}^\eps}) = 0 \label{eq:PFSS_HB_cc}
\end{equation}\\

\noindent \textit{Expression for calibrated $\de$}: Following a large number of calibration simulations using the single-edge notched tension test, the following approximate functional form was fitted to the calibrated $\de$ values (expression valid for both regular and compression corrected versions of the theory)
\begin{equation}
    \de \approx \frac{0.115}{\mcf^2}\left(\frac{\lch^\texttt{HB}}{\eps}\right) + \frac{1}{\mcf} , \quad \lch^\texttt{HB} = \frac{3 E G_c}{8 {\stsHB}^2} = \frac{3G_c}{16\WtsHB} \label{eq:de_fitted_HB}
\end{equation}
where $\lch^\texttt{HB}$ is the Irwin characteristic length and $\mcf$ is the mesh correction factor defined in \cref{eq:mcf}. The expression is valid in the range $0.5 \lesssim {\lch^\texttt{HB}}/{\eps} \lesssim 10$.

\subsection{Mogi-Coulomb}\label{subsec:Mogi_surface}

Another widely used strength criterion in rock mechanics is the Mogi-Coulomb criterion that 
 reduces exactly to the Mohr-Coulomb criterion for loadings where there is no distinct intermediate principal stress ($\smax=\sint$ or $\sint=\smin$) but makes use of all 3 principal stresses in general loadings, allowing it to better capture polyaxial strength data \cite{al2005relation,al2006stability}. Thus, the Mogi-Coulomb criterion enjoys a similar advantage as the 3D Hoek-Brown criterion in that it includes the effect of intermediate principal stress, but it lacks the advantage of the Hoek-Brown criterion in being able to capture the nonlinear relationship between axial stress $s_1$ and confining pressure $s_3$ in conventional triaxial compression tests. 
  The criterion can be written as follows (with $\vect{\beta} = [\beta_1,\beta_2]$)
    \begin{equation}
 \boxed{  \mathcal{F}_{\mathtt{MgC} }\equiv {F}_{\texttt{MgC}}({\bm{\sigma}}; \vect{\beta})  = \beta_1 \smTwo + \beta_2 \tauoct - 1 = 0 }\label{eq:generalform_MgC} 
\end{equation}
with
\begin{equation}
    \beta_1 = \frac{1}{\scsMC} - \frac{1}{\stsMC}, \quad \beta_2 = \frac{3\sqrt{2}}{4}\left(\frac{1}{\stsMC} + \frac{1}{\scsMC}\right), \label{eq:betas_MgC}
\end{equation}
where the uniaxial tensile strength $\stsMC$ and the uniaxial compressive strength $\scsMC$ are the same as that for the Mohr-Coulomb surface. The fitted parameters and curve to the conventional triaxial compression strength data of Indiana Limestone are unchanged from the Mohr-Coulomb criterion.\\
 
\noindent \textit{Mg-C surface incorporation into phase-field theory: } If the uniaxial tensile and compressive strengths are chosen as the  calibration locations on the strength surface, the solution for the parameters $\vect{\Delta \beta^\eps}$  can be calculated using \cref{eq:deltabetaeps_sol}  as
\begin{equation}
    \Delta \beta^\eps_1 = \frac{2}{\omega_\eps}\left(\frac{\WtsMC}{\stsMC} - \frac{\WcsMC}{\scsMC}\right), \quad \Delta \beta^\eps_2 = -\frac{3\sqrt{2}}{2 \omega_\eps}\left(\frac{\WtsMC}{\stsMC} + \frac{\WcsMC}{\scsMC}\right) \label{eq:Dbe_MgC}
\end{equation}
The corresponding phase-field strength surface can be written using \cref{eq:First_phase_ss_general} as 
\begin{equation}
 {\mathcal{F}}^\eps_\mathtt{MgC} \equiv   \frac{2 W({\bm{\sigma}})}{\omega_\eps} + {F}_\mathtt{MgC}({\bm{\sigma}}; \vect{\beta} + \vect{\Delta \beta^\eps}) = 0 \label{eq:PFSS_MgC}
\end{equation}
where $\vect{\beta}$ and $\Dbe$ are given in \cref{eq:betas_MgC,eq:Dbe_MgC}. For the compression corrected solution, the parameters $\Dbecc$  can be calculated using \cref{eq:deltabetaeps_sol_I1corrected}  as
\begin{equation}
    \Delta \beta^\eps_{\texttt{cc}_1} = \frac{2\WtsMC}{\omega_\eps\stsMC}, \quad \Delta \beta^\eps_{\texttt{cc}_2} = -\frac{3\sqrt{2}\WtsMC}{2 \omega_\eps \stsMC} \label{eq:Dbe_MgC_cc}
\end{equation}
and the phase-field strength surface from \eqref{eq:phase_ss_general_I1} as
\begin{equation}
 {\mathcal{F}}^\eps_\mathtt{MgC, cc} \equiv   \frac{2 W({\bm{\sigma}})}{\omega_\eps}f_\texttt{cc}(I_1) + {F}_\mathtt{MgC}({\bm{\sigma}}; \vect{\beta} + \vect{\Delta \beta_\mathtt{cc}^\eps}) = 0 \label{eq:PFSS_MgC_cc}
\end{equation}\\

\noindent \textit{Expression for calibrated $\de$}: Following a large number of calibration simulations using the single-edge notched tension test, the following approximate functional form was fitted to the calibrated $\de$ values for the compression corrected version of the theory,
\begin{equation}
    \de \approx \frac{0.14}{\mcf^2}\left(\frac{\lch^\texttt{MgC}}{\eps}\right) + \frac{0.96}{\mcf}, \quad \lch^\texttt{MgC} = \lch^\texttt{MC}~~(\text{defined in \cref{eq:de_fitted_MC}}) \label{eq:de_fitted_MgC}
\end{equation}
where $\lch^\texttt{MgC}$ is the Irwin characteristic length and $\mcf$ is the mesh correction factor defined in \cref{eq:mcf}. The expression is valid in the range $0.5 \lesssim {\lch^\texttt{MgC}}/{\eps} \lesssim 10$.\\

\section{Analysis and results}\label{Sec: Numerics}

In this section, we demonstrate the ability of the strength-incorporated phase-field theory to capture fracture nucleation and propagation across various regimes. We begin by modeling strength-controlled fracture nucleation in the absence of pre-existing cracks under uniform multiaxial stress loadings in \Cref{subsec:StrengthLimit}. We then demonstrate fracture toughness-controlled propagation of large cracks in accordance with Griffith's theory by modeling the double cantilever beam test in \Cref{subsec:DCBT}. Finally, we demonstrate the ability of the theory to capture fracture nucleation due to mediation between strength and fracture toughness in \Cref{subsec:Interplay}. Specifically, in \Cref{sec:SENT} we demonstrate this mediation in the single-edge notched tension test  and in \Cref{sec:3PBT} for the single-edge notched bend test. For the sake of brevity, results for the Mogi-Coulomb surface will only be presented for the multiaxial stress loading problem in \Cref{subsec:StrengthLimit} (since the M-C and Mg-C results for the other problems are nearly identical).\\

\noindent \textit{Material parameters for Indiana Limestone: } As mentioned earlier, we will be using Indiana Limestone as our model material here for all our analysis and results. The material strength parameters calibrated in the previous section are summarized below,
\begin{equation}
    \stsMC = 17.5 \text{MPa},\quad \scsMC = 50 \text{MPa},\quad \stsHB = 12.3 \text{MPa},\quad \scsHB =   50 \text{MPa}
\end{equation}
The other elastic and fracture parameters are adopted from \cite{epp2018laboratory,schmidt1976fracture},  
\begin{equation}
    E = 30~\text{GPa},\quad \nu = 0.23,\quad G_c = 30~\text{J}/\text{m}^2
\end{equation}
The associated Irwin characteristic lengths for the Mohr-Coulomb, Hoek-Brown and Mogi-Coulomb strength surfaces defined in \cref{eq:de_fitted_MC,eq:de_fitted_HB,eq:de_fitted_MgC} are calculated to be $\lch^\texttt{MC} \approx 1.1$ mm, $\lch^\texttt{HB} \approx 2.23$ mm, and $\lch^\texttt{MgC} \approx 1.1$ mm, respectively. Values of the parameter $\de$ were calibrated using the single-edge notched tension test (described in \Cref{sec:SENT}) and have been tabulated in \ref{app:de_values}. These calibrated values were used in part to fit the approximate functional forms provided in the previous section. The finite element simulations to follow were carried out using the open-source finite element platform FEniCS.

\subsection{The strength limit: fracture nucleation under uniform multi-axial stress}
\label{subsec:StrengthLimit}

\begin{figure}[h]
    \centering   \includegraphics[width=0.37\textwidth]{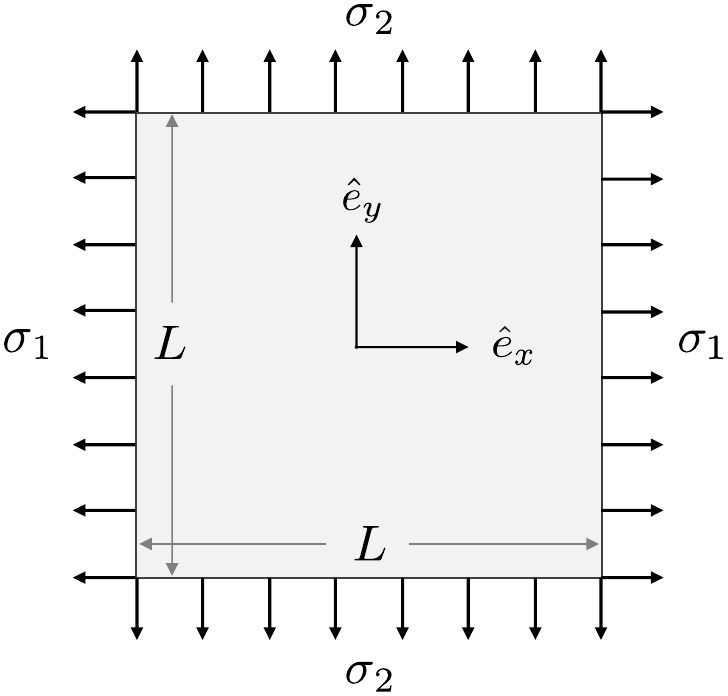}
    \caption{Illustration of the boundary value problem used to model strength-controlled failure under uniform multiaxial stress.}
    \label{fig:PFSS_illustration}
\end{figure}

We begin by considering strength-controlled fracture nucleation when a pristine block of material with no pre-existing defects is subject to a spatially uniform stress state $\bm\sigma$ by application of tractions $\tilde{\textbf{t}} = \bm{\sigma} \nten{N}$ at the boundaries. Consider $\bm\sigma = |\bm\sigma| \hat{\bm\sigma}$ where the magnitude of the stress is given by $|\bm\sigma| = \sqrt{\bm\sigma : \bm\sigma}$ and $\hat{\bm\sigma}$ is a unit direction in stress space. For fixed $\hat{\bm\sigma}$ and increasing $|\bm\sigma|$, the material behaves elastically with $v=1$ everywhere until a critical value of $|\bm\sigma|$ is reached when the material starts fracturing ($v<1$). The critical stress value defines the location on the strength surface corresponding to the loading direction $\hat{\bm\sigma}$; the collection of critical stress values for all possible $\hat{\bm\sigma}$ defines the strength surface of the material.\\

In this section, we plot the strength surface predicted by phase-field theory for various regularization lengths and compare it with the material's strength surface. The strength surfaces predicted by the regular and compression-corrected versions of the phase-field theory are plotted by solving the analytical equations (\ref{eq:First_phase_ss_general}) and (\ref{eq:phase_ss_general_I1}), respectively, as well as by running finite element simulations for both. For simplicity, we consider a plane stress setting where $\sigma_3 = 0$ and $\sigma_1$ and $\sigma_2$ are the principal stresses in the plane of loading. For the finite element simulations,  we consider a square block of material sample with dimension {$L=50~\lch$} such that the face normals $\pm\hat{e}_x$ and $\pm\hat{e}_y$ are aligned with the direction of principal stresses $\sigma_1$ and $\sigma_2$ respectively (see \Cref{fig:PFSS_illustration}). Thus the faces are loaded by the uniform tractions $\tilde{\textbf{t}} = \pm \sigma_1~\hat{e}_x$ and $\tilde{\textbf{t}} = \pm \sigma_2~\hat{e}_y$ as shown in \Cref{fig:PFSS_illustration}. A uniform triangular mesh with characteristic size $h=\eps/5$ is chosen to discretize the simulation geometry. The critical stress at fracture nucleation is numerically evaluated by checking for when the condition $v\approx0$ is satisfied anywhere
in the specimen which is also seen to correspond closely to the stress value at which $v<1$ anywhere since the phase field loses stability at the critical stress\footnote{Stress-controlled simulations are unstable at the critical stress and in some cases, the numerical solver fails to converge at the onset of failure. In such cases, the stress at solver failure is taken as the critical stress.}. The critical stress is then plotted for various directions in the principal stress space, where any given direction is defined by a fixed ratio between the principal stresses $\sigma_1$ and $\sigma_2$. \\

\begin{figure}[ht]
    \centering   \includegraphics[width=\textwidth]{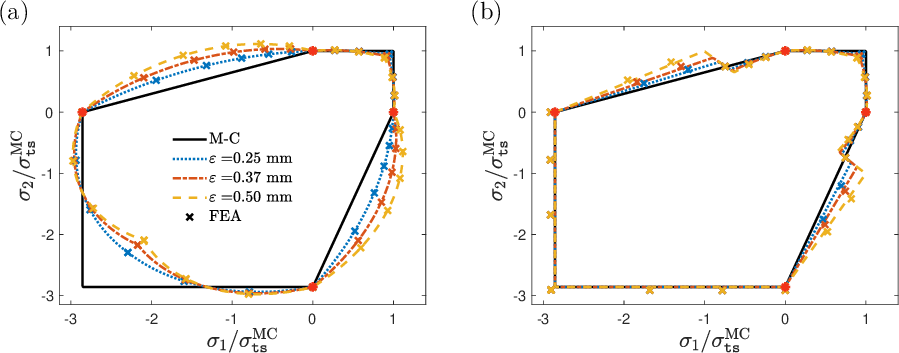}
    \caption{Comparison of plots of material strength surface and phase-field strength surface for the Mohr-Coulomb criterion in principal stress space under plane stress conditions using material parameters of Indiana Limestone. Solid black line is the exact Mohr-Coulomb material strength surface $\mathcal{F}_{\mathtt{MC} }$ defined in \cref{eq:generalform_MC}. (a) Phase-field strength surface ${\mathcal{F}}^\eps_\mathtt{MC}$ defined in \cref{eq:PFSS_MC}. (b) Compression corrected phase-field strength surface ${\mathcal{F}}^\eps_\mathtt{MC, cc}$ defined in \cref{eq:PFSS_MC_cc}.  The critical stresses from finite element analysis are plotted as crosses in the corresponding colors for the different $\varepsilon$.  The legend is the same for both plots. The red star-marked points are the calibrated strength locations (uniaxial tensile strength $\stsMC$ and uniaxial compressive strength $\scsMC$) that are exactly captured by the phase-field theory for all $\varepsilon$.}
    \label{fig:MCPFSS}
\end{figure}

We first plot the results for the Mohr-Coulomb criterion in \Cref{fig:MCPFSS}. \Cref{fig:MCPFSS}(a) plots the results for the regular version of the theory, while \Cref{fig:MCPFSS}(b) plots the results for the compression corrected version. The exact Mohr-Coulomb material strength surface $\mathcal{F}_{\mathtt{MC} }$ as defined by \cref{eq:generalform_MC} is also shown with a solid black line. The strength surfaces predicted by the phase-field theory are shown for three different values of the regularization length (0.25 \text{mm}, 0.37 \text{mm}, 0.50 \text{mm}) which correspond to a range of about $23-45\%$ of the Irwin characteristic length $\lch^\texttt{MC}$ (within typical range employed in numerical simulations). The colored lines represent the analytically predicted phase-field strength surfaces (\cref{eq:PFSS_MC} for ${\mathcal{F}}^\eps_\mathtt{MC}$ in \Cref{fig:MCPFSS}(a) and \cref{eq:PFSS_MC_cc} for ${\mathcal{F}}^\eps_\mathtt{MC, cc}$ in \Cref{fig:MCPFSS}(b)) whereas the colored crosses represent the results of the finite element simulations, which agree closely with the analytical predictions. The red star-marked points are the chosen strength calibration locations that are exactly captured at all $\eps$, specifically the uniaxial tensile strength $\stsMC$ and uniaxial compressive strength $\scsMC$ here. See \cite{chockalingam2025construction} for other interesting choices of strength calibrations.\\

It is seen in \Cref{fig:MCPFSS} that the phase-field strength surface exactly matches the true material strength surface at the strength calibration locations, whereas the mismatch at other stress states decreases for decreasing value of regularization length $\eps$. Note that the phase-field strength surface is mathematically guaranteed to converge to the material strength surface in the sharp limit of $\eps \searrow 0$. However it is seen from \Cref{fig:MCPFSS}(a) that the phase-field approximation can be quite off from the true material strength at finite practical values of regularization length in the compressive regime ($\sigma_1 + \sigma_2 < 0)$. As expected, the compression corrected version of the theory improves the strength prediction in this regime, as seen in \Cref{fig:MCPFSS}(b).\\

\begin{figure}[h]
    \centering   \includegraphics[width=\textwidth]{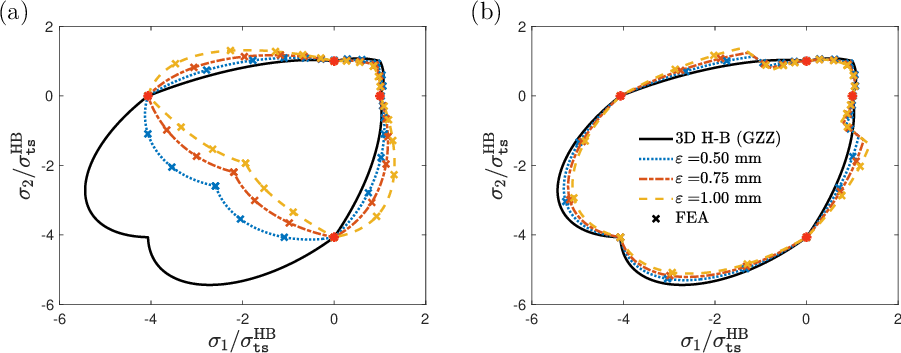}
    \caption{Comparison of plots of material strength surface and phase-field strength surface for the 3D Hoek-Brown criterion in principal stress space under plane stress conditions using material parameters of Indiana Limestone. Solid black line is the exact 3D Hoek-Brown material strength surface $\mathcal{F}_{\mathtt{HB} }$ defined in \cref{eq:generalform_HB}. (a) Phase-field strength surface ${\mathcal{F}}^\eps_\mathtt{HB}$ defined in \cref{eq:PFSS_HB}. (b) Compression corrected phase-field strength surface ${\mathcal{F}}^\eps_\mathtt{HB, cc}$ defined in \cref{eq:PFSS_HB_cc}.  The critical stresses from finite element analysis are plotted as crosses in the corresponding colors for the different $\varepsilon$.  The legend is the same for both plots. The red star-marked points are the calibrated strength locations (uniaxial tensile strength $\stsHB$ and uniaxial compressive strength $\scsHB$) that are exactly captured by the phase-field theory for all $\varepsilon$.}
    \label{fig:HBPFSS}
\end{figure}

\begin{figure}[h]
    \centering   \includegraphics[width=\textwidth]{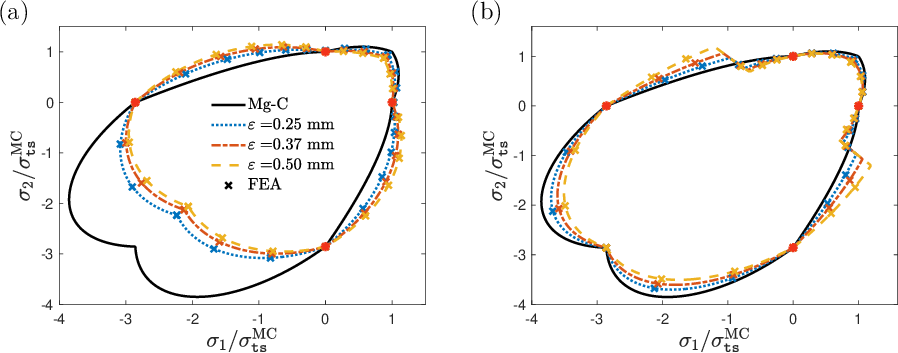}
    \caption{Comparison of material strength surface and phase-field strength surface for the Mogi-Coulomb criterion in principal stress space under plane stress conditions, plotted for different values of the regularization length $\eps$, using material parameters of Indiana Limestone. Solid black line is the exact Mogi-Coulomb material strength surface $\mathcal{F}_{\mathtt{MgC} }$ defined in \cref{eq:generalform_MgC}. (a) Phase-field strength surface ${\mathcal{F}}^\eps_\mathtt{MgC}$ defined in \cref{eq:PFSS_MgC}. (b) Compression corrected strength surface ${\mathcal{F}}^\eps_\mathtt{MgC, cc}$ defined in \cref{eq:PFSS_MgC_cc}.  The critical stresses from finite element analysis are plotted as crosses in the corresponding colors for the different $\varepsilon$.  The legend is the same for both plots. The red star-marked points are the calibrated strength locations (uniaxial tensile strength $\stsMC$ and uniaxial compressive strength $\scsMC$) that are exactly captured by the phase-field theory for all $\varepsilon$. }
    \label{fig:Mogi}
\end{figure}

Next, we plot the results for the 3D Hoek-Brown criterion in \Cref{fig:HBPFSS}; \Cref{fig:HBPFSS}(a) plots the results for the regular version of the theory while \Cref{fig:HBPFSS}(b) plots the results for the compression corrected version. The solid black line is the exact 3D Hoek-Brown material strength surface $\mathcal{F}_{\mathtt{HB}}$ defined in \cref{eq:generalform_HB}.  The phase-field results are now demonstrated for regularization length values of 0.5 mm, 0.75 mm, and 1 mm, which again correspond to a range of about $23-45\%$ of the Irwin characteristic length $\lch^\texttt{HB}$. The phase-field strength surface for the regular version of the theory ${\mathcal{F}}^\eps_\mathtt{HB}$ is defined in \cref{eq:PFSS_HB}, whereas that for the compression corrected version ${\mathcal{F}}^\eps_\mathtt{HB, cc}$ is defined in \cref{eq:PFSS_HB_cc}. All the observations made about the results of the Mohr-Coulomb criterion continue to hold for the 3D Hoek-Brown criterion as well. Namely, (i) the phase-field strength surface converges to the exact material strength surface as $\varepsilon \searrow 0$, (ii) the phase-field strength prediction is exact at chosen strength calibration locations irrespective of $\varepsilon$, and (iii) compression correction enhances strength predictions for compressive loading states. Similar observations also hold for the results of the Mogi-Coulomb surface, which are plotted in \Cref{fig:Mogi} (see caption for relevant equations).\\

{Due to its better approximation of the material strength surface, we will employ the compression corrected version of the theory for our numerical demonstrations in the coming sections, but we note that the regular version works equally well for tension-dominated tests.}

\subsection{The Griffith limit: fracture nucleation and propagation from a large crack}
\label{subsec:DCBT}

\begin{figure}[h]
    \centering   \includegraphics[width=\textwidth]{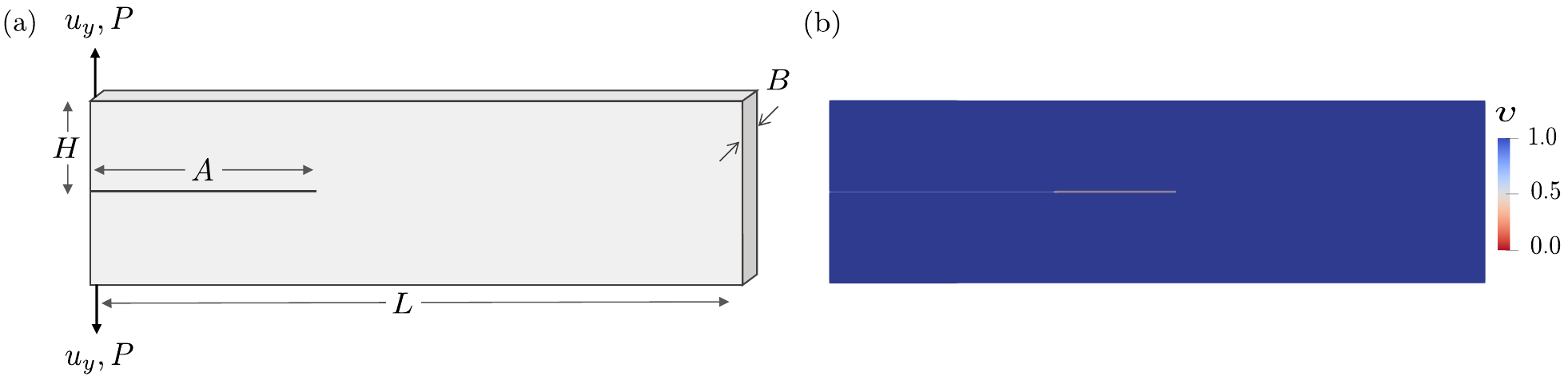}
    \caption{The double cantilever beam test (DCBT) for large crack propagation. (a) Illustration of the test. (b) The crack visualized at an applied displacement $u_y = 2 u_{y,\text{crit}}$ through a plot of the phase field variable $v$ from simulations using the Mohr-Coulomb strength-incorporated phase-field theory and $\varepsilon = 0.5$ mm. The red region represents the grown crack.}
    \label{fig:DCBTIllus}
\end{figure}

We now demonstrate the ability of the theory to capture the Griffith physics of nucleation and propagation from large pre-existing cracks controlled by fracture toughness,  using the double cantilever beam test (DCBT). The double cantilever beam test \cite{gilman1960direct,gillis1964double,mostovoy1967use,wiederhorn1968critical} is a standard configuration used to study Mode I fracture propagation under controlled conditions.
It was recently identified as one of several benchmark problems that a complete fracture model should be able to capture accurately \cite{KAMAREI2026118449}. In the displacement-controlled version of this test, a pre-cracked specimen with two symmetrical arms is loaded in opening mode through applied displacement ($u_y$) at the beam tips, promoting controlled crack propagation. See \Cref{fig:DCBTIllus}(a) for an illustration of the test and the associated specimen dimensions---span $L$, half-thickness $H$, initial crack length $A$, and out-of-plane width $B$. The response load $P$ and the crack length $a$ are concurrently measured as a function of the applied displacement to obtain load-displacement ($P-u_y$) and crack growth ($a-u_y$) response.\\

The elastic analysis and energy release rate expressions for the double cantilever geometry were provided by Gross and Srawley \cite{gross1966stress} for $L>A+3H$. Combined with the Griffith criticality condition in \cref{Griffith}, the analytical solution for the crack growth and load-displacement response can be shown to be as follows for monotonically increasing applied displacement $u_y$ and plane stress conditions,
\begin{equation}
\textbf{Crack growth response:}\quad
\begin{cases}
\textit{Elastic branch:} &
a = A,
\quad
u_y < u_{y,\text{crit}}, \\[8pt]
\textit{Fracture branch:} &
u_y = 
\dfrac{2}{\sqrt{3}}
\sqrt{\dfrac{G_c H}{E}}
\dfrac{a^2}{H^2}
\dfrac{g_1(a/H)}{g_2(a/H)},
\quad
u_y > u_{y,\text{crit}},
\end{cases}
\label{eq:crack_growth_response}
\end{equation}
where 
\begin{equation}
    u_{y,\text{crit}} = \dfrac{2}{\sqrt{3}} \sqrt{\dfrac{G_c H}{E}} \dfrac{A^2}{H^2} \dfrac{g_1(A/H)}{g_2(A/H)} 
\end{equation}
and
\begin{equation}
    g_1(y) = 1 + \frac{\gamma \sqrt{3}}{2 y} +\frac{\gamma^2}{4 y^2},\quad g_2(y) = 1 + \frac{\gamma}{2\sqrt{3} y}, \quad \gamma = 2.38
\end{equation}

\begin{equation}
\textbf{Load--displacement response:}\quad
\begin{cases}
\textit{Elastic branch:} &
\dfrac{P}{B} = 
\dfrac{E H^3}{4 A^3}
\dfrac{u_y}{g_1(A/H)},
\quad
u_y < u_{y,\text{crit}}, \\[10pt]
\textit{Fracture branch:} &
\dfrac{P}{B} =
\sqrt{\dfrac{G_c E H}{12}}
\dfrac{H}{a}
\dfrac{1}{g_2(a/H)},
\quad
u_y > u_{y,\text{crit}},\\
& \text{where}~ a = f(u_y)~ \text{from \cref{eq:crack_growth_response}}
\end{cases}
\label{eq:load_disp_response}
\end{equation}
\\

\begin{figure}[h]
    \centering   \includegraphics[width=\textwidth]{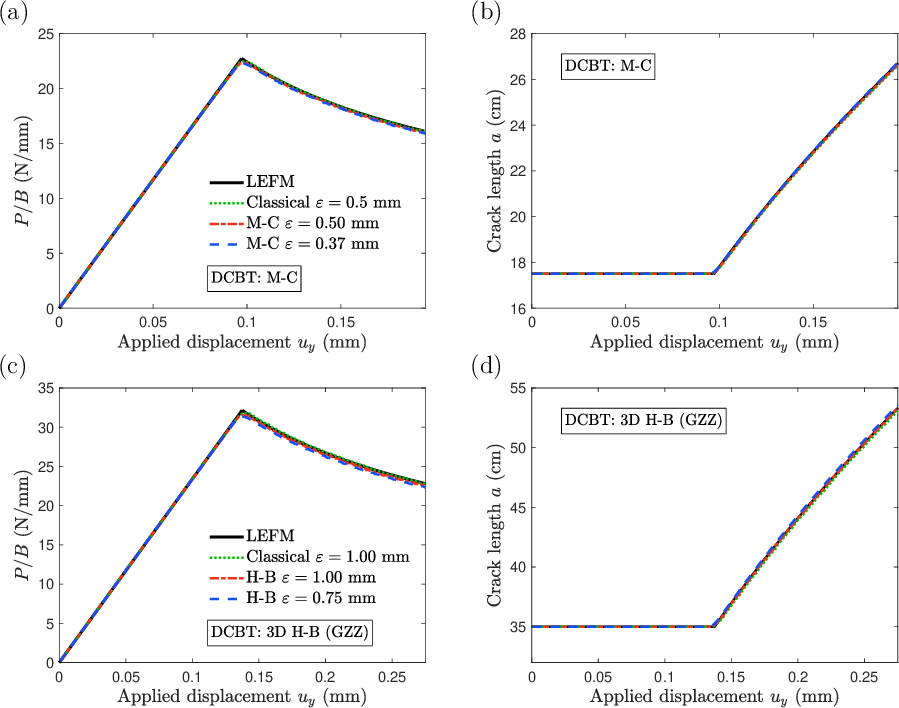}
    \caption{Results for the double cantilever beam test (DCBT) using Indiana Limestone parameters. The solid black line is the theoretical response described in equations (\ref{eq:crack_growth_response})-(\ref{eq:load_disp_response}) using linear elastic fracture mechanics (LEFM). Plots of load per unit out-of-plane width ($P/B$) as a function of applied displacement $u_y$ for the (a) M-C surface and (c) 3D H-B surface.    Plots of crack length $a$ as a function of applied displacement $u_y$ for the (b) M-C surface and (d) 3D H-B surface. (a) and (b) share the same legend, and so do (c) and (d). }
    \label{fig:DCBTresult}
\end{figure}

The finite element simulations for DCBT were carried out in a plane-stress setting with a geometrical initial crack and ran for monotonically increasing $u_y$ up to a value of $2 u_{y,\text{crit}}$. The specimen dimensions were chosen based on the following factors: (i) The beam compliance should be small so that the deformations are not large enough to induce geometric nonlinearity---accordingly we choose a small value of $A/H = 2.5$ (since the elastic compliance scales nearly cubically with $A/H$ as seen from \cref{eq:load_disp_response}), (ii) the structural response needs to be dominated by large crack Griffith physics and thus we require $A,H,L >> \lch$, and (iii) the length of the beam $L$ needs to be greater than $a+3H$ at the largest crack length $a$ encountered in the simulations so that the energy release rate expression used to derive the analytical solution remains valid. Accordingly, the following specimen dimensions were chosen for the simulations
\begin{align}
&\text{Dimensions for M-C surface simulations} : \quad A = 17.5~\text{cm},\quad H = 7~\text{cm},\quad  L = 50.4~\text{cm} \\
&\text{Dimensions for 3D H-B surface simulations} : \quad A = 35~\text{cm},\quad H = 14~\text{cm},\quad  L = 100.8~\text{cm},  
\end{align}
We note that the dimensions for the 3D H-B surface simulations are double those for the M-C surface simulations since the Irwin characteristic length is nearly twice for the 3D H-B surface. The evolution of the crack length was monitored by tracking the farthest-advanced location of $v=0$ near the symmetry line. We set $v=0$ in a tiny region around the front of the initial crack---this is referred to as ``damaged notch conditions'' and is necessary to avoid overestimation of critical stresses based on Griffith criticality condition in the phase-field approach \cite{Tanne18}. An unstructured triangular mesh was used that is highly refined near the crack path with a characteristic mesh size $h = \eps/5$ and is coarse away from the crack path. The exact location of the displacement application on the beam arms does not affect the results much, but for the results shown here, it is applied across the entire loaded face of the beam. This was seen to give the best match to the expressions of Gross and Srawley \cite{gross1966stress} when running a purely elastic simulation. The load was calculated through an integral of the residuals \cite{bathe2006finite} and was seen to match well with the integral of the tractions on the beam arms.\\

The analytical and numerical results for DCBT are plotted in \Cref{fig:DCBTresult} using Indiana Limestone parameters. The load-displacement response ($P/B$ vs $u_y$) for the case with the M-C surface is plotted in \Cref{fig:DCBTresult}(a), and that for the 3D H-B surface is plotted in \Cref{fig:DCBTresult}(c). The crack growth response ($a$ vs $u_y$) is plotted for the M-C surface in \Cref{fig:DCBTresult}(b) and for the 3D H-B surface in \Cref{fig:DCBTresult}(d). The analytical curves are plotted as black lines using equations (\ref{eq:crack_growth_response})-(\ref{eq:load_disp_response}) and are labeled as LEFM (linear elastic fracture mechanics). The load-displacement curves linearly increase with the applied displacement in the elastic branch until the critical displacement is reached, following which the load reduces with increasing displacement as the crack length increases in the fracture branch. The numerical results are plotted for both the strength-incorporated and classical phase-field theories (the latter obtained by setting $c_e = 0$ and $\de = 1$). Results for different values of the regularization length are shown for the strength-incorporated theory. The classical phase-field theory is expected to capture Griffith's large-crack physics and serves as a baseline for validation. It is seen in \Cref{fig:DCBTresult} that the results for both the classical and strength-incorporated phase-field theories show excellent match with the analytical curves. \textit{Thus the strength-incorporated phase-field theory with the M-C, H-B, and Mg-C (not shown here but verified) surfaces 
is able to preserve the ability of the classical phase-field theory to capture Griffith physics in the large crack limit while also being able to capture strength physics in the absence of cracks as shown in the previous section (which the classical phase-field theory cannot \cite{lopez2025classical,kamarei2025nucleation})}. A snapshot of the phase field variable from the numerical simulations is shown at $u_y = 2 u_{y,\text{crit}}$ for the M-C surface case (with $\eps = 0.5$ mm) in \Cref{fig:DCBTIllus}(b) to visualize the grown crack (red region).\\

As far as the authors are aware, this is the first accurate reproduction of the analytical DCBT results using phase-field theory reported in the literature. The simulations are intricate and several careful numerical considerations went into producing the above numerical results---a couple major ones are highlighted here for the sake of future practitioners: (i) Numerical phase-field simulations often add a tiny residual stiffness for numerical stability in fractured regions with $v=0$, such an addition can adversely affect the DCBT simulation results, (ii) Often, symmetric simulations are performed by exploiting the geometric symmetry of the problem, such as about the crack plane in the DCBT problem. However, the use of such symmetry introduces a non-existent flux condition for the phase field at the symmetry plane that can potentially affect the accuracy of simulations. Thus, it is preferable to model the full geometry of the problem. 

\begin{remark}
   \rm Note that the values of $\de$ used in the double cantilever beam simulations above were calibrated by matching simulation predictions to known analytical results for the critical stress in a single-edge notched tension test (described in \Cref{sec:SENT}), which is a completely different fracture test. However, the same value of $\de$ captures the critical stress and fracture response exactly in the double cantilever beam test, so it is independent of the boundary-value problem. While this observation has previously been made for the case of Drucker-Prager strength surface \cite{KBFLP20, KRLP22, KLDLP24, KKLP24, LK24, KAMAREI2026118449}, the results here suggest universal applicability to general strength surfaces.
\end{remark}

\subsection{Strength-Griffith mediated fracture nucleation}
\label{subsec:Interplay}

We have demonstrated that the strength-incorporated phase-field theory accurately captures both strength-dominated behavior in the absence of cracks and fracture-toughness-dominated (Griffith) behavior in the presence of large cracks. We now use this theory to demonstrate fracture nucleation governed by an interplay between strength and toughness physics, and to show that it smoothly transitions between the two limiting regimes. This is achieved by progressively varying the initial crack length from small to large values and showing that the critical stress for fracture nucleation evolves continuously from the strength-based prediction to the Griffith prediction.\\

We first illustrate this behavior for the single-edge-notched tension test (\Cref{sec:SENT}), where the stress field approaches spatial uniformity in the strength limit of vanishing crack length. This configuration has previously been used to demonstrate the capability of the strength-incorporated phase-field theory to capture the interplay between strength and toughness for the Drucker-Prager strength surface \cite{KBFLP20,KAMAREI2026118449}. We then examine the same transition in the single-edge notched bend test (\Cref{sec:3PBT}), a geometry where the stress field remains spatially non-uniform even in the limit of vanishing crack length---representing the first demonstration of this interplay for any strength surface.

\subsubsection{Uniform stress in the vanishing crack limit: The single-edge notched tension test}
\label{sec:SENT}

\begin{figure}[h]
    \centering   \includegraphics[width=0.25\textwidth]{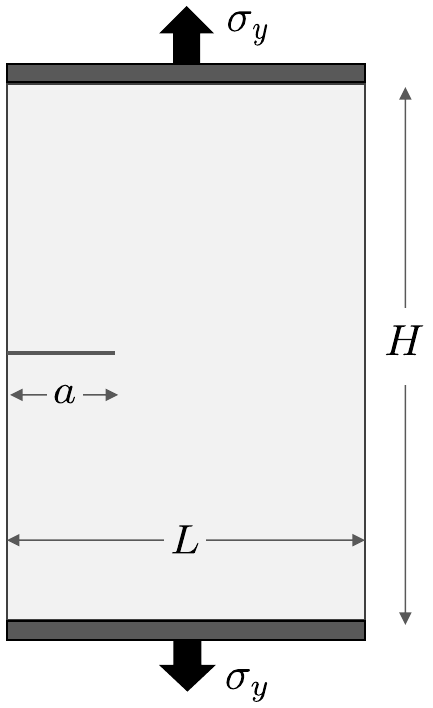}
    \caption{Illustration of the single-edge notched tension (SENT) test. }
    \label{fig:SENTILLUS}
\end{figure}

The single-edge notched tension (SENT) test is a standard test used to study Mode I fracture. The specimen consists of a rectangular plate with a pre-crack of length $a$ introduced from one edge and the ends of the plate parallel to the crack are clamped or gripped and pulled apart in tension, as shown in \Cref{fig:SENTILLUS}. We consider load-controlled testing, and apply a uniform stress  $\sigma_y$ at the loaded ends. In the strength limit of vanishing crack lengths (small $a/\lch$), the specimen undergoes uniform tensile stress and thus the critical stress for fracture nucleation ($\sigma_y^\text{crit}$) is simply the uniaxial tensile strength of the material. In the Griffith limit of large crack lengths (large $a/\lch$), expressions for the energy release rate for the SENT test \cite{tada1973stress} can be combined with the Griffith criticality condition in \cref{Griffith} to obtain a critical fracture nucleation stress (note that the crack unstably propagates beyond the critical stress in load-controlled testing). The analytical critical fracture nucleation stress can be written as follows
\begin{equation}
\rm{\textbf{Critical stress:}}\quad \begin{cases}
\textit{Strength limit (small } a/l_{\text{ch}}): &
F(\boldsymbol{\sigma}; \vect{\beta}) = 0
\;\Rightarrow\;
\sigma_y^\text{crit} = \sts, \\[10pt]
\textit{Griffith limit (large } a/l_{\text{ch}}): &
G = G_c
\;\Rightarrow\;
\sigma_y^\text{crit} = f_{\texttt{ST}}(a/L)\,
\sqrt{\dfrac{E G_c}{\pi a}}.
\end{cases}
\label{eq:failure_condition}
\end{equation}
with
\begin{equation}
f_{\texttt{ST}}(a/L) =
\dfrac{
\cos\!\left(\dfrac{\pi a}{2L}\right)
}{
\left[
0.752
+ 2.02\,\dfrac{a}{L}
+ 0.37\,\Big(1 - \sin\!\left(\dfrac{\pi a}{2L}\right)\Big)^{3}
\right]
\sqrt{\dfrac{2L}{\pi a}\tan\!\left(\dfrac{\pi a}{2L}\right)}}.
\label{eq:h_def}
\end{equation}
\\

Finite element simulations are once again carried out in plane stress conditions and using an unstructured triangular mesh that is highly refined near the crack with a characteristic mesh size $h = \text{min}(\eps,a)/5$. The crack length $a$ is varied between $0.01~\lch$ to $60~\lch$ and the following sample dimensions are chosen (see \Cref{fig:SENTILLUS})
\begin{equation}
    H = 600~\lch, \quad L = 200~\lch 
\end{equation}
Simulations are run using both undamaged and damaged notch conditions, where in the latter, the phase field $v$ is set to zero in a small region near the crack tip. As mentioned in the previous section, damaged notch conditions are necessary in the phase-field method to avoid overestimation of critical stresses in the Griffith limit. The numerical critical stress for fracture nucleation is evaluated by checking when $v\approx0$ at the crack tip in the undamaged notch case and slightly ahead of the pre-damaged zone for damaged notch conditions\footnote{{Load-controlled simulations are unstable at the critical stress and the numerical solver can fail at the onset of failure. 
In such cases, the stress at solver failure is taken as the critical stress.}}. We note that shortly after critical stress is reached, the crack propagates catastrophically under load-controlled simulations.\\

\begin{figure}[h]
    \centering   \includegraphics[width=\textwidth]{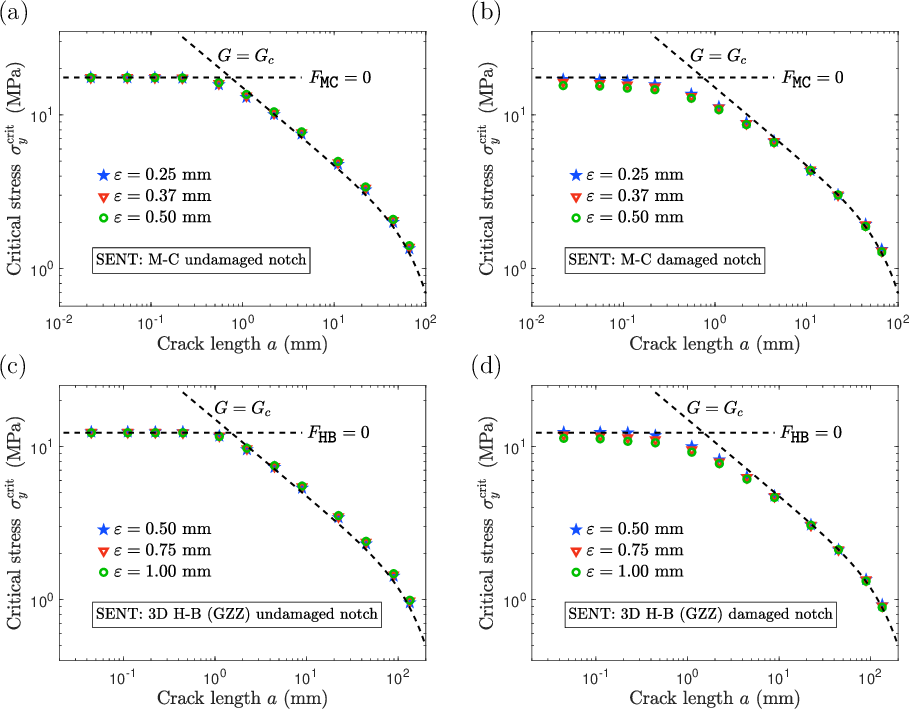}
    \caption{Plot of the critical stress as a function of crack length for SENT tests using Indiana Limestone parameters.  M-C surface results shown in (a) for undamaged notch conditions and in (b) for damaged notch conditions. 3D H-B surface results shown in (c) for undamaged notch conditions and in (d) for damaged notch conditions. Black dashed lines are the analytical solution for strength and Griffith limits from \cref{eq:failure_condition}. Numerical results from finite element analysis are shown using colored markers for different regularization lengths. }
    \label{fig:SENT_result}
\end{figure}

The analytical and numerical estimates of the critical stress $\sigma_y$ for the SENT test are plotted as a function of the crack length $a$ in \Cref{fig:SENT_result}.  The analytical critical stresses based on the strength and Griffith criterion are plotted using black dashed lines. The lower of the two stresses governs fracture nucleation, and thus it is seen that the Griffith-based prediction governs large crack lengths, whereas the strength-based prediction governs small crack lengths, as expected. The numerical results are shown for both M-C and 3D H-B strength surfaces, as well as damaged and undamaged notch conditions for each. In all cases, the numerical phase-field predictions closely match the analytical strength and Griffith limits, and smoothly transition between the two. Thus the ability of the strength-incorporated phase-field theory to model the interplay of strength and toughness physics has been demonstrated.\\

The critical stress in the large-crack Griffith limit matches more closely with the analytical prediction when damaged-notch conditions are imposed, whereas undamaged-notch conditions result in a slight over-prediction, consistent with established observations \cite{Tanne18}. Conversely, the strength limit is better predicted in undamaged notch conditions, whereas damaged notch conditions lead to under-prediction of critical stress. This is because the addition of a small fractured zone weakens the material; the strength limit predictions for damaged notch conditions improve for smaller $\eps$ since the size of the region over which $v=0$ was imposed at the crack tip was scaled with $\eps$. The strength limit predictions for undamaged notch conditions match exactly with the theoretical curve, irrespective of $\eps$, since the uniaxial tensile strength is a calibrated strength location in the theory. Note that in real application settings with no pre-existing large cracks, strength-based fracture nucleation and subsequent Griffith propagation of phase field cracks would be captured accurately without the need for ``damaged notch conditions'' that were required here for geometric cracks.

\subsubsection{Non-uniform stress in the vanishing crack limit: The single-edge notched bend test}
\label{sec:3PBT}

\begin{figure}[h]
    \centering   \includegraphics[width=0.6\textwidth]{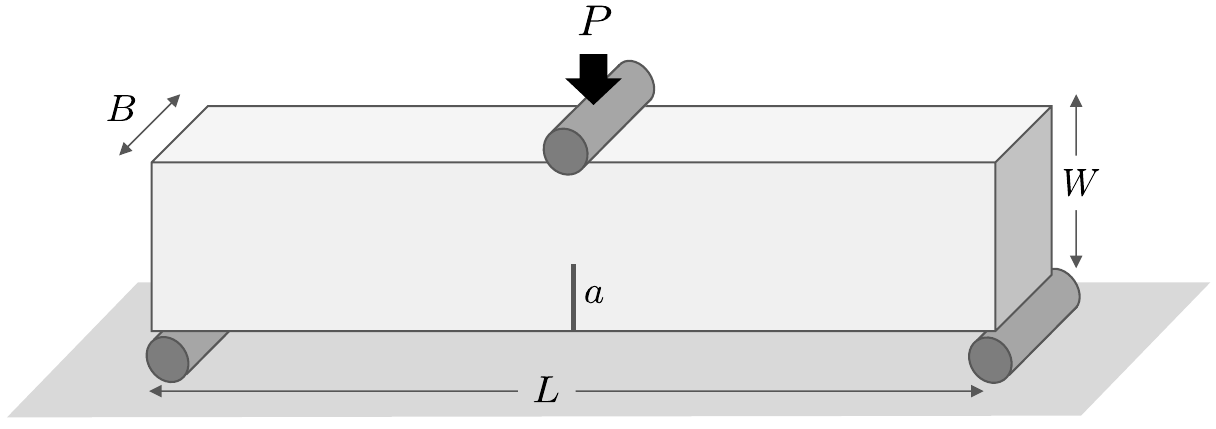}
    \caption{Illustration of the single-edge notched bend test for fracture. }
    \label{fig:3PB_Illus}
\end{figure}

We now model the single-edge notched bend (SENB) test (also  called the 3-point bend test), which is one of the commonly used fracture tests in rocks. The test specimen typically consists of a thick rectangular beam with a single-edge pre-crack (of length $a$), simply supported at its end with cylindrical rollers and loaded at its midspan with a cylindrical roller contact, as illustrated in \Cref{fig:3PB_Illus}. The rollers distribute the applied load $P$ and reaction force at supports over a small finite width.  Since the beams are thick in the out-of-plane direction (large $B$ in \Cref{fig:3PB_Illus}), plane strain conditions prevail generally. We consider load-controlled testing where the applied force is $P$ and investigate the critical load $P^\text{crit}$ that causes fracture nucleation at the crack tip.\\

The elastic stress state in the limit of vanishing crack length ($a/\lch \to 0$) is non-uniform for the SENB test and can be approximated using an analytical beam theory (Euler-Bernoulli or Timoshenko) solution assuming the applied load $P$ is a line load spanning $B$ at the midspan, the supports are line-supports spanning $B$ at the ends, and that plane strain conditions prevail. Fracture nucleation in this limit can be expected to occur at the crack tip when the stress-state there first satisfies the strength condition in \cref{eq:SSurf-0}. Accordingly, the critical load for fracture nucleation in the strength limit of vanishing crack length for the SENB test under plane strain conditions can be approximated as follows (using either Euler–Bernoulli or Timoshenko beam theory, since both predict the same stress state at the midspan on the bottom surface of the uncracked beam) 
\begin{align}
\frac{P^\text{crit}}{B} &= 
\frac{2W^2}{3L}\,\sts^\texttt{pe},
\quad
\text{where }
F(\bm{\sigma}_\texttt{ts}^\texttt{pe}; \vect{\beta}) = 0, \quad \boldsymbol{\sigma}_\texttt{ts}^\texttt{pe} =
\sts^\texttt{pe}\,\hat{\mathbf{e}}_1 \otimes \hat{\mathbf{e}}_1
+ \nu\,\sts^\texttt{pe}\,\hat{\mathbf{e}}_3 \otimes \hat{\mathbf{e}}_3,
\label{eq:Pcrit_strength}
\end{align}
and $\{\hat{\mathbf{e}}_1, \hat{\mathbf{e}}_2, \hat{\mathbf{e}}_3\}$ form a constant orthonormal Cartesian basis. Note that for the M-C surface, $\sts^\texttt{pe} = \sts (= \stsMC$), since the M-C criterion ignores the intermediate principal stress $\sigint =  \nu\,\sts^\texttt{pe}$. In the Griffith limit of large crack lengths (large $a/\lch$), expressions for the energy release rate for the SENB test \cite{tada1973stress} can be combined with the Griffith criticality condition in \cref{Griffith} to obtain a critical fracture nucleation load (the crack unstably propagates beyond this load). When $L=4W$ (analytical energy release rate expression is valid for this specific dimension ratio, and we will use the same in our simulations), the critical fracture nucleation load based on Griffith condition under plane strain conditions can be written as (assuming line load at midspan and line supports)
\begin{align}
\frac{P^{\text{crit}}}{B} = 
\frac{2W^2}{3L} \sqrt{\frac{E G_c}{(1-\nu^2)\pi a}}\frac{1}{f_\texttt{SB}(a/W)},\quad f_\texttt{SB}(\alpha) = 
\frac{1}{\sqrt{\pi}}
\frac{
1.99 - \alpha(1 - \alpha)\!\left(2.15 - 3.93\,\alpha + 2.7\,\alpha^2\right)
}{
(1 + 2\alpha)(1 - \alpha)^{3/2} \label{eq:Pcrit_Griffith}
}
\end{align}\\

Finite element simulations are run in plane strain conditions and using an unstructured triangular mesh that is highly refined near the crack with characteristic mesh size $h = \text{min}(\eps,a)/5$. The crack length $a$ is varied between $0.03~\lch$ to $30~\lch$ and the following sample dimensions are chosen (see \Cref{fig:3PB_Illus})
\begin{equation}
    W \approx 40~\lch, \quad L = 4W
\end{equation}
The widths of the supports were taken to be $0.001W$ each, and the width over which the load is applied was taken to be $0.001W$. Simulations are run using both undamaged and damaged notch conditions, where in the latter, the phase field $v$ is set to zero in a small region near the crack tip. The numerical critical load for fracture nucleation is evaluated by checking when $v \approx 0$ at the crack tip in the undamaged notch case and slightly ahead of the pre-damaged zone for damaged notch conditions\footnote{{Load-controlled simulations are unstable at the critical load and the numerical solver can fail at the onset of failure. 
In such cases, the load at solver failure is taken as the critical load.}}. We note that shortly after the critical load is reached, the crack propagates catastrophically.\\

\begin{figure}[h]
    \centering   \includegraphics[width=\textwidth]{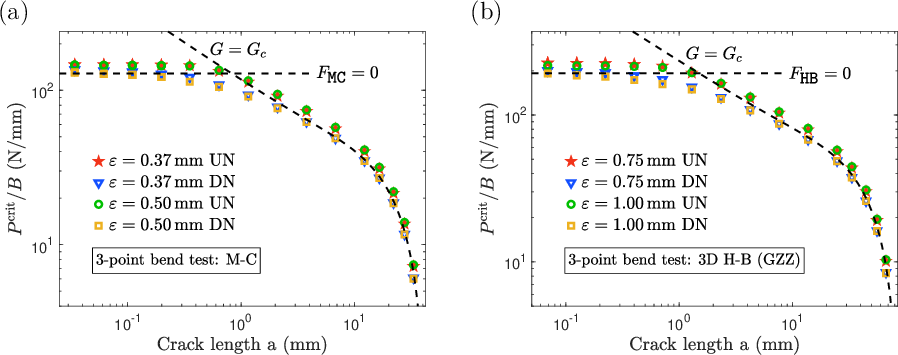}
    \caption{Plot of the critical stress as a function of crack length for SENB tests using Indiana Limestone parameters. Results for (a) M-C surface and (b) 3D H-B surface shown for undamaged notch (UN) and damaged notch (DN) conditions. Black dashed lines are the analytical solution for strength and Griffith limits from \cref{eq:Pcrit_strength,eq:Pcrit_Griffith}, respectively. Note that the strength limit curve is an approximation using beam theory. Numerical results from finite element analysis are shown using colored markers for different regularization lengths. }
    \label{fig:3PB_result}
\end{figure}

The analytical and numerical estimates of the critical load for the SENB test are plotted as a function of the crack length $a$ in \Cref{fig:3PB_result}.  The analytical critical loads based on the strength and Griffith criterion are plotted using black dashed lines. The lower of the two critical loads governs fracture nucleation, and thus it is seen that the Griffith-based prediction governs large crack lengths, whereas the strength-based prediction governs small crack lengths, as expected. The numerical results are shown for both M-C and 3D H-B strength surfaces, as well as damaged and undamaged notch conditions for each. In all cases, the numerical phase-field predictions agree reasonably well with analytical strength and Griffith limits and smoothly transition between the two. \\

 Once again, the critical load in the large-crack Griffith limit matches more closely with the analytical prediction when damaged-notch conditions are imposed, whereas undamaged-notch conditions result in a slight over-prediction. However, in the strength limit, the numerical results are seen to be higher than the analytical estimates for undamaged notch conditions or for smaller regularization lengths in the damaged notch conditions. This could potentially be due to the approximate nature of the analytical estimate (since it uses beam theory, line load assumption etc.), the non-uniformity of the stress, or due to numerical difficulties such as localized damage from high stresses near the load, which lowers the stresses in the beam compared to elastic beam theory predictions for a given applied load $P$. Nevertheless, the overall trends in critical load match well with the analytical expectations. Thus, the ability of the strength-incorporated phase-field theory to model the interplay of strength and toughness physics has been demonstrated in a problem setting with non-uniform stresses, even in the absence of cracks. We have thus demonstrated the ability of the theory to model fracture in all regimes - the strength regime, the Griffith regime and the intermediate regime with an interplay of strength and Griffith physics.

\section{Summary and final comments}\label{Sec: Final Comments}

Employing the driving-force solution presented by Chockalingam (2025) \cite{chockalingam2025construction}, the present work incorporated the Mohr--Coulomb (M-C), 3D Hoek--Brown (3D H-B), and Mogi--Coulomb (Mg-C) strength surfaces into the phase-field theory proposed by Kumar et al. (2020) \cite{KBFLP20}. Through finite element simulations of several canonical fracture problems with known analytical solutions, the phase-field theory was comprehensively validated across diverse fracture regimes. Key takeaways are summarized below.
\begin{itemize}
    \item All previous implementations of the strength-incorporated phase-field theory were restricted to the Drucker--Prager (D-P) surface. This work reports the first implementation of alternative, more general strength surfaces. In particular, the 3D H-B surface represents the first incorporation of a criterion whose strength function is nonlinear (non-homogeneous) in stress.
    \item The strength predicted by the phase-field theory converges to the material's strength surface in the limit of vanishing regularization length $\eps \searrow 0$, while exactly reproducing the calibrated strength locations at all $\varepsilon$. The compression-corrected formulation substantially improves strength predictions in compressive loading states for finite $\eps$.
    \item The theory retains the toughness-based Griffith criticality condition for large cracks (demonstrated through simulations of the double-cantilever-beam test) while simultaneously predicting strength-controlled nucleation for pristine samples under arbitrary multiaxial loadings (demonstrated through uniform-stress loading).
    \item Consistent with observations from prior studies for the D-P surface, the calibrated correction parameter $\delta_\varepsilon$ is independent of the boundary-value problem for the surfaces considered here as well. This points to the potential universality of this property---a key requirement for the theory to be meaningfully valuable for analyzing general problems---across arbitrary strength surfaces.
    \item The theory captures the interplay and yields smooth transitions between strength- and toughness-dominated regimes, as shown by analyzing the single-edge-notched tension and single-edge-notched bend tests for varying crack lengths. The latter represents the first demonstration of this effect in a problem where the stress field is non-uniform in the uncracked limit, demonstrating the robustness of the theory for complex fracture scenarios.
\end{itemize}

Although Indiana Limestone was chosen as a benchmark material for the numerical demonstration, identical trends were observed across a wide range of material parameters. Thus, the strength surfaces presented here enable modeling fracture in a broad class of materials, including rocks, ceramics, concrete, ice, and bone. The results further provide confidence in the applicability of the proposed driving-force formulation to brittle materials with arbitrary strength surfaces. Finally, this framework lays the foundation for extending the phase-field theory beyond brittle fracture to capture fracture nucleation and propagation in ductile materials, where the nucleation criteria can be more complex \cite{BaiWierzbicki2010,keralavarma2016criterion,keralavarma2020ductile} and coupled to constitutive physics such as plasticity, porosity, and microstructure evolution. The present work thus represents a key step toward advancing the phase-field approach into a truly general theory of fracture.

\FloatBarrier

\appendix
\renewcommand\thesection{Appendix~\Alph{section}}

\section{Calibrated values of $\de$ for Indiana Limestone}
\label{app:de_values}

Here we present values of the $\de$ parameter calibrated specifically for the Indiana Limestone parameters used in this manuscript (which are used in part for the approximate fitted functional forms provided in \Cref{subsec:strength_surfaces}) when $h=\eps/5$ where $h$ is the characteristic mesh size in the fracturing region. The calibrations were performed by matching the critical stress predictions from the single-edge notched tension test (described in \Cref{sec:SENT}) to their known analytical expressions from LEFM. Results for the M-C, 3D H-B, and Mg-C surfaces are tabulated in \Cref{tab:de_MC,tab:de_HB,tab:de_MgC}, respectively. 

\begin{table}[h!]
  \centering
  \caption{Calibrated $\de$ values for Mohr-Coulomb strength surface and Indiana Limestone parameters.}
  \begin{tabular}{lccc}
    \toprule
    Theory & $\eps = 0.25$ mm & $\eps = 0.37$ mm & $\eps = 0.50$ mm \\
    \midrule
    Regular \& Compression-corrected & 1.779 & 1.471 & 1.293 \\
    \bottomrule
  \end{tabular}
  \label{tab:de_MC}
\end{table}

\begin{table}[h!]
  \centering
  \caption{Calibrated $\de$ values for 3D Hoek-Brown strength surface and Indiana Limestone parameters.}
  \begin{tabular}{lcccc}
    \toprule
    Theory & $\eps = 0.50$ mm & $\eps = 0.75$ mm & $\eps = 1.00$ mm \\
    \midrule
    Regular & 1.365 & 1.227 & 1.169 \\
    Compression-corrected & 1.368 & 1.212 & 1.134 \\
    \bottomrule
  \end{tabular}
  \label{tab:de_HB}
\end{table}

\begin{table}[h!]
  \centering
  \caption{Calibrated $\de$ values for Mogi-Coulomb strength surface and Indiana Limestone parameters.}
  \begin{tabular}{lcccc}
    \toprule
    Theory & $\eps = 0.25$ mm & $\eps = 0.37$ mm & $\eps = 0.50$ mm \\
    \midrule
    Regular & 1.344 & 1.147 & 1.061 \\
    Compression-corrected & 1.455 & 1.270 & 1.185 \\
    \bottomrule
  \end{tabular}
  \label{tab:de_MgC}
\end{table}
\FloatBarrier

\section*{Acknowledgements}

\noindent  SC would like to acknowledge the Lillian Gilbreth Postdoctoral Fellowship at Purdue University. ABT gratefully acknowledges support from the U.S. Army Research Office under Award W911NF-24-1-0244. AK acknowledges support from the National Science Foundation, United States, through the grant CMMI-2404808.
\bibliographystyle{elsarticle-num-names}
\bibliography{refs_combined}

@article{mostovoy1967use,
  title={Use of crack-line-loaded specimens for measuring plane-strain fracture toughness},
  author={Mostovoy, Sheldon},
  journal={Journal of materials},
  volume={2},
  number={3},
  pages={661--681},
  year={1967}
}

@article{rosendahl2019equivalent,
  title={Equivalent strain failure criterion for multiaxially loaded incompressible hyperelastic elastomers},
  author={Rosendahl, PL and Drass, M and Felger, J and Schneider, J and Becker, W},
  journal={International Journal of Solids and Structures},
  volume={166},
  pages={32--46},
  year={2019},
  publisher={Elsevier}
}

@incollection{labuz2014mohr,
  title={Mohr--Coulomb failure criterion},
  author={Labuz, Joseph F and Zang, Arno},
  booktitle={The ISRM Suggested Methods for Rock Characterization, Testing and Monitoring: 2007-2014},
  pages={227--231},
  year={2014},
  publisher={Springer}
}

@article{gilman1960direct,
  title={Direct measurements of the surface energies of crystals},
  author={Gilman, John J},
  journal={Journal of applied physics},
  volume={31},
  number={12},
  pages={2208--2218},
  year={1960},
  publisher={American Institute of Physics}
}

@article{tada1973stress,
  title={The stress analysis of cracks},
  author={Tada, Hiroshi and Paris, Paul C and Irwin, George R},
  journal={Handbook, Del Research Corporation},
  volume={34},
  number={1973},
  year={1973}
}

@article{gillis1964double,
  title={Double-cantilever cleavage mode of crack propagation},
  author={Gillis, Peter P and Gilman, John Joseph},
  journal={Journal of Applied Physics},
  volume={35},
  number={3},
  pages={647--658},
  year={1964}
}

@article{wiederhorn1968critical,
  title={Critical analysis of the theory of the double cantilever method of measuring fracture-surface energies},
  author={Wiederhorn, SM and Shorb, AM and Moses, RL},
  journal={Journal of Applied Physics},
  volume={39},
  number={3},
  pages={1569--1572},
  year={1968},
  publisher={American Institute of Physics}
}

@article{zhang2008generalized,
  title={A generalized three-dimensional Hoek--Brown strength criterion},
  author={Zhang, L},
  journal={Rock mechanics and rock engineering},
  volume={41},
  number={6},
  pages={893--915},
  year={2008},
  publisher={Springer}
}

@article{al2005relation,
  title={Relation between the Mogi and the Coulomb failure criteria},
  author={Al-Ajmi, Adel M and Zimmerman, Robert W},
  journal={International Journal of Rock Mechanics and Mining Sciences},
  volume={42},
  number={3},
  pages={431--439},
  year={2005},
  publisher={Elsevier}
}

@article{KAMAREI2026118449,
title = {Nine circles of elastic brittle fracture: A series of challenge problems to assess fracture models},
journal = {Computer Methods in Applied Mechanics and Engineering},
volume = {448},
pages = {118449},
year = {2026},
issn = {0045-7825},
author = {Farhad Kamarei and Bo Zeng and John E. Dolbow and Oscar Lopez-Pamies},
}

@techreport{gross1966stress,
  title={Stress-intensity factors by boundary collocation for single-edge-notch specimens subject to splitting forces},
  author={Gross, Bernard and Srawley, John E.},
  institution={National Aeronautics and Space Administration},
  type={NASA Technical Note D-2395},
  year={1966},
  address={Washington, DC}
}

@article{al2006stability,
  title={Stability analysis of vertical boreholes using the Mogi--Coulomb failure criterion},
  author={Al-Ajmi, Adel M and Zimmerman, Robert W},
  journal={International journal of rock mechanics and mining sciences},
  volume={43},
  number={8},
  pages={1200--1211},
  year={2006},
  publisher={Elsevier}
}

@article{ghimire2022determination,
  title={Determination of Mohr--Coulomb failure envelope, mechanical properties and UPV of commercial cement-lime mortar},
  author={Ghimire, Amrit and Noor-E-Khuda, Sarkar and Ullah, Shah Neyamat and Suntharavadivel, Thuraichamy},
  journal={Materials and Structures},
  volume={55},
  number={4},
  pages={111},
  year={2022},
  publisher={Springer}
}

@article{schmidt1976fracture,
  title={Fracture-toughness testing of limestone: KIc of indiana limestone was measured using three-point-bend specimens, and toughness is seen to increase with crack length much like many aluminum alloys},
  author={Schmidt, Richard A},
  journal={Experimental mechanics},
  volume={16},
  number={5},
  pages={161--167},
  year={1976},
  publisher={Springer}
}

@article{lelovic2020determination,
  title={Determination of Mohr-coulomb parameters for modelling of concrete},
  author={Lelovic, Selimir and Vasovic, Dejan},
  journal={Crystals},
  volume={10},
  number={9},
  pages={808},
  year={2020},
  publisher={MDPI}
}

@mastersthesis{epp2018laboratory,
  title={Laboratory rock strength measurements of saturated carbonates: Implications for the Grosmont Formation},
  author={Epp, Tyson R.},
  school={University of Alberta},
  address={Edmonton, Canada},
  year={2018}
}

@article{li2024stress,
  title={Stress-dependent Mohr--Coulomb shear strength parameters for intact rock},
  author={Li, Hao and Pel, Leo and You, Zhenjiang and Smeulders, David},
  journal={Scientific Reports},
  volume={14},
  number={1},
  pages={17454},
  year={2024},
  publisher={Nature Publishing Group UK London}
}

@inproceedings{HoekBrown1988,
  author       = {Hoek, Evert and Brown, E. T.},
  title        = {The Hoek–Brown Failure Criterion – A 1988 Update},
  booktitle    = {Proceedings of the 15th Canadian Rock Mechanics Symposium},
  editor       = {Curran, J. H.},
  location     = {Toronto, Ontario, Canada},
  organization = {Civil Engineering Department, University of Toronto},
  pages        = {31--38},
  year         = {1988}
}

@article{hoek1983strength,
  title={Strength of jointed rock masses},
  author={Hoek, Evert},
  journal={Geotechnique},
  volume={33},
  number={3},
  pages={187--223},
  year={1983},
  publisher={Thomas Telford Ltd}
}

@article{wang2008identification,
  title={Identification of material parameters based on Mohr--Coulomb failure criterion for bisphosphonate treated canine vertebral cancellous bone},
  author={Wang, Xiang and Allen, Matthew R and Burr, David B and Lavernia, Enrique J and Jeremi{\'c}, Boris and Fyhrie, David P},
  journal={Bone},
  volume={43},
  number={4},
  pages={775--780},
  year={2008},
  publisher={Elsevier}
}

@article{arenson2005mathematical,
  title={Mathematical descriptions for the behaviour of ice-rich frozen soils at temperatures close to 0 C},
  author={Arenson, Lukas U and Springman, Sarah M},
  journal={Canadian Geotechnical Journal},
  volume={42},
  number={2},
  pages={431--442},
  year={2005},
  publisher={NRC Research Press Ottawa, Canada}
}

@article{hoek2002hoek,
  title={Hoek-Brown failure criterion-2002 edition},
  author={Hoek, Evert and Carranza-Torres, Carlos and Corkum, Brent and others},
  journal={Proceedings of NARMS-Tac},
  volume={1},
  number={1},
  pages={267--273},
  year={2002}
}

@article{zhang2007three,
  title={Three-dimensional Hoek-Brown strength criterion for rocks},
  author={Zhang, Lianyang and Zhu, Hehua},
  journal={Journal of Geotechnical and Geoenvironmental Engineering},
  volume={133},
  number={9},
  pages={1128--1135},
  year={2007},
  publisher={American Society of Civil Engineers}
}

@inproceedings{hoek1992modified,
  title={A modified Hoek--Brown failure criterion for jointed rock masses},
  author={Hoek, E and Wood, D and Shah, S},
  booktitle={Rock Characterization: ISRM Symposium, Eurock'92, Chester, UK, 14--17 September 1992},
  pages={209--214},
  year={1992},
  organization={Thomas Telford Publishing}
}

@article{murrell1963criterion,
  title={A criterion for brittle fracture of rocks and concrete under triaxial stress and the effect of pore pressure on the criterion},
  author={Murrell, SAF},
  journal={Rock mechanics},
  pages={563--577},
  year={1963},
  publisher={Pergamon Oxford}
}

@article{hoek1980empirical,
  title={Empirical strength criterion for rock masses},
  author={Hoek, Evert and Brown, Edwin T},
  journal={Journal of the geotechnical engineering division},
  volume={106},
  number={9},
  pages={1013--1035},
  year={1980},
  publisher={American Society of Civil Engineers}
}

@article{singh2011modified,
  title={Modified Mohr--Coulomb criterion for non-linear triaxial and polyaxial strength of intact rocks},
  author={Singh, Mahendra and Raj, Anil and Singh, Bhawani},
  journal={International Journal of Rock Mechanics and Mining Sciences},
  volume={48},
  number={4},
  pages={546--555},
  year={2011},
  publisher={Elsevier}
}

@article{handin1967effects,
  title={Effects of the intermediate principal stress on the failure of limestone, dolomite, and glass at different temperatures and strain rates},
  author={Handin, John and Heard, HC and and Magouirk, JN},
  journal={Journal of geophysical research},
  volume={72},
  number={2},
  pages={611--640},
  year={1967},
  publisher={Wiley Online Library}
}

@inproceedings{hoskins1969failure,
  title={The failure of thick-walled hollow cylinders of isotropic rock},
  author={Hoskins, ER},
  booktitle={International Journal of Rock Mechanics and Mining Sciences \& Geomechanics Abstracts},
  volume={6},
  number={1},
  pages={99--125},
  year={1969},
  organization={Elsevier}
}

@article{mogi1971fracture,
  title={Fracture and flow of rocks under high triaxial compression},
  author={Mogi, Kiyoo},
  journal={Journal of Geophysical Research},
  volume={76},
  number={5},
  pages={1255--1269},
  year={1971},
  publisher={Wiley Online Library}
}

@inproceedings{takahashi1989effect,
  title={Effect of the intermediate principal stress on strength and deformation behavior of sedimentary rocks at the depth shallower than 2000 m},
  author={Takahashi, M and Koide, H},
  booktitle={ISRM international symposium},
  pages={ISRM--IS},
  year={1989},
  organization={ISRM}
}

@article{haimson2000new,
  title={A new true triaxial cell for testing mechanical properties of rock, and its use to determine rock strength and deformability of Westerly granite},
  author={Haimson, B and Chang, Chan-dong},
  journal={International Journal of Rock Mechanics and Mining Sciences},
  volume={37},
  number={1-2},
  pages={285--296},
  year={2000},
  publisher={Elsevier}
}

@article{haimson2002true,
  title={True triaxial strength of the KTB amphibolite under borehole wall conditions and its use to estimate the maximum horizontal in situ stress},
  author={Haimson, Bezalel C and Chang, Chandong},
  journal={Journal of Geophysical Research: Solid Earth},
  volume={107},
  number={B10},
  pages={ETG--15},
  year={2002},
  publisher={Wiley Online Library}
}

@article{chockalingam2025construction,
  title={On the Construction of Explicit Analytical Driving Forces for Crack Nucleation in the Phase Field Approach to Brittle Fracture With Application to Mohr--Coulomb and Drucker--Prager Strength Surfaces},
  author={Chockalingam, S},
  journal={Journal of Applied Mechanics},
  volume={92},
  number={4},
  pages={041002},
  year={2025},
  publisher={American Society of Mechanical Engineers}
}

@article{lopez2025classical,
  title={Classical variational phase-field models cannot predict fracture nucleation},
  author={Lopez-Pamies, Oscar and Dolbow, John E and Francfort, Gilles A and Larsen, Christopher J},
  journal={Computer Methods in Applied Mechanics and Engineering},
  volume={433},
  pages={117520},
  year={2025},
  publisher={Elsevier}
}

@article{KLDLP24,
  title={The strength of the Brazilian fracture test},
  author={Kumar, Aditya and Liu, Yangyuanchen and Dolbow, John E and Lopez-Pamies, Oscar},
  journal={Journal of the Mechanics and Physics of Solids},
  volume={182},
  pages={105473},
  year={2024},
  publisher={Elsevier}
}

@article{keralavarma2016criterion,
  title={A criterion for void coalescence in anisotropic ductile materials},
  author={Keralavarma, SM and Chockalingam, S},
  journal={International Journal of Plasticity},
  volume={82},
  pages={159--176},
  year={2016},
  publisher={Elsevier}
}

@article{keralavarma2020ductile,
  title={Ductile failure as a constitutive instability in porous plastic solids},
  author={Keralavarma, SM and Reddi, D and Benzerga, AA},
  journal={Journal of the Mechanics and Physics of Solids},
  volume={139},
  pages={103917},
  year={2020},
  publisher={Elsevier}
}

@article{kamarei2025nucleation,
  title={Nucleation of fracture: The first-octant evidence against classical variational phase-field models},
  author={Kamarei, Farhad and Dolbow, John E and Lopez-Pamies, Oscar},
  journal={Journal of Applied Mechanics},
  volume={92},
  number={1},
  pages={014502},
  year={2025},
  publisher={American Society of Mechanical Engineers}
}

@article{KBFLP20,
	title={Revisiting nucleation in the phase-field approach to brittle fracture},
	author={Kumar, Aditya and Bourdin, Blaise and Francfort, Gilles A and Lopez-Pamies, Oscar},
	journal={Journal of the Mechanics and Physics of Solids},
	volume={142},
	pages={104027},
	year={2020},
	publisher={Elsevier}
}

@article{KFLP18,
	title={Fracture and healing of elastomers: A phase-transition theory and numerical implementation},
	author={Kumar, Aditya and Francfort, Gilles A and Lopez-Pamies, Oscar},
	journal={Journal of the Mechanics and Physics of Solids},
	volume={112},
	pages={523--551},
	year={2018},
	publisher={Elsevier}
}

@article{KLP20,
	title={The phase-field approach to self-healable fracture of elastomers: A model accounting for fracture nucleation at large, with application to a class of conspicuous experiments},
	author={Kumar, Aditya and Lopez-Pamies, Oscar},
	journal={Theoretical and Applied Fracture Mechanics},
	volume={107},
	pages={102550},
	year={2020},
	publisher={Elsevier}
}

@article{KRLP22,
	title={The revisited phase-field approach to brittle fracture: application to indentation and notch problems},
	author={Kumar, Aditya and Ravi-Chandar, K and Lopez-Pamies, Oscar},
	journal={International Journal of Fracture},
	volume={237},
	number={1-2},
	pages={83--100},
	year={2022},
	publisher={Springer}
}

@article{KLP21,
	title={The poker-chip experiments of Gent and Lindley (1959) explained},
	author={Kumar, Aditya and Lopez-Pamies, Oscar},
	journal={Journal of the Mechanics and Physics of Solids},
	volume={150},
	pages={104359},
	year={2021},
	publisher={Elsevier}
}

@book{bathe2006finite,
  title={Finite element procedures},
  author={Bathe, Klaus-J{\"u}rgen},
  year={2006},
  publisher={Klaus-Jurgen Bathe}
}

@article{Bourdin00,
	title={Numerical experiments in revisited brittle fracture},
	author={Bourdin, Blaise and Francfort, Gilles A and Marigo, Jean-Jacques},
	journal={Journal of the Mechanics and Physics of Solids},
	volume={48},
	number={4},
	pages={797--826},
	year={2000},
	publisher={Elsevier}
}

@article{Bourdin08,
	title={The variational approach to fracture},
	author={Bourdin, Blaise and Francfort, Gilles A and Marigo, Jean-Jacques},
	journal={Journal of elasticity},
	volume={91},
	pages={5--148},
	year={2008},
	publisher={Springer}
}

@article{Tanne18,
	title={Crack nucleation in variational phase-field models of brittle fracture},
	author={Tann{\'e}, Erwan and Li, Tianyi and Bourdin, Blaise and Marigo, J-J and Maurini, Corrado},
	journal={Journal of the Mechanics and Physics of Solids},
	volume={110},
	pages={80--99},
	year={2018},
	publisher={Elsevier}
}

@article{Francfort98,
	title={Revisiting brittle fracture as an energy minimization problem},
	author={Francfort, Gilles A and Marigo, J-J},
	journal={Journal of the Mechanics and Physics of Solids},
	volume={46},
	number={8},
	pages={1319--1342},
	year={1998},
	publisher={Elsevier}
}

@article{KKLP24,
  title={The poker-chip experiments of synthetic elastomers explained},
  author={Kamarei, Farhad and Kumar, Aditya and Lopez-Pamies, Oscar},
  journal={Journal of the Mechanics and Physics of Solids},
  pages={105683},
  year={2024},
  publisher={Elsevier}
}

@article{BaiWierzbicki2010,
  title={Application of extended Mohr--Coulomb criterion to ductile fracture},
  author={Bai, Yuanli and Wierzbicki, Tomasz},
  journal={International journal of fracture},
  volume={161},
  number={1},
  pages={1--20},
  year={2010},
  publisher={Springer}
}

@article{LK24,
title = {Emergence of tension–compression asymmetry from a complete phase-field approach to brittle fracture},
journal = {International Journal of Solids and Structures},
volume = {309},
pages = {113170},
year = {2025},
author = {Chang Liu and Aditya Kumar}
}

@article{consoli2014mohr,
  title={Mohr--Coulomb failure envelopes of lime-treated soils},
  author={Consoli, NC and Da Silva Lopes Jr, L and Consoli, BS and Festugato, L},
  journal={G{\'e}otechnique},
  volume={64},
  number={2},
  pages={165--170},
  year={2014},
  publisher={Thomas Telford Ltd}
}

@article{anand2025fracture,
  title={Fracture of rock-like materials: A gradient-damage theory},
  author={Anand, Lallit},
  journal={International Journal of Solids and Structures},
  pages={113739},
  year={2025},
  publisher={Elsevier}
}

@article{KDK2025Comparison,
title = {A comparison of phase field models of brittle fracture incorporating strength, I:Mixed-mode loading},
journal = {Engineering Fracture Mechanics},
pages = {111679},
year = {2025},
issn = {0013-7944},
author = {Umar Khayaz and Aarosh Dahal and Aditya Kumar}
}

@article{clayton2025MC,
  title={Modeling ice cliff stability using a new Mohr--Coulomb-based phase field fracture model},
  author={Clayton, Theo and Duddu, Ravindra and Hageman, Tim and Mart{\'\i}nez-Pa{\~n}eda, Emilio},
  journal={Journal of Glaciology},
  volume={71},
  pages={e70},
  year={2025},
  publisher={Cambridge University Press}
}

@article{chang2018application,
  title={Application of the Mohr-Coulomb yield criterion for rocks with multiple joint sets using fast Lagrangian analysis of continua 2D (FLAC2D) software},
  author={Chang, Lifu and Konietzky, Heinz},
  journal={Energies},
  volume={11},
  number={3},
  pages={614},
  year={2018},
  publisher={MDPI}
}

@article{niwa1967mortarMC,
  title={Failure criterion of cement mortar under triaxial compression},
  author={NIWA, Yoshiji and KOBAYASHI, Shoichi},
  journal={Memoirs of the Faculty of Engineering, Kyoto University},
  volume={29},
  number={1},
  pages={1--15},
  year={1967},
  publisher={Faculty of Engineering, Kyoto University}
}

@article{mamot2018iceMC,
  title={A temperature-and stress-controlled failure criterion for ice-filled permafrost rock joints},
  author={Mamot, Philipp and Weber, Samuel and Schr{\"o}der, Tanja and Krautblatter, Michael},
  journal={The Cryosphere},
  volume={12},
  number={10},
  pages={3333--3353},
  year={2018},
  publisher={Copernicus GmbH}
}

\end{document}